\documentclass[fleqn,usenatbib]{mnras}

\usepackage{newtxtext,newtxmath}

\usepackage[T1]{fontenc}
\usepackage{ae,aecompl}

\usepackage{bm}
\usepackage{graphicx} 
\usepackage{amsmath} 
\usepackage{xcolor} 

\newcommand{\g}{\mathrm{g}}
\newcommand{\dd}{\mathrm{d}}

\let\vec\bm

\title[SPH for multigrain dust]{A smoothed particle hydrodynamics algorithm for
multigrain dust with separate sets of particles}

\author[Mentiplay, Price, Pinte, and Laibe]{%
   \parbox{\textwidth}{%
      Daniel Mentiplay\(^{1}\)\thanks{daniel.mentiplay@monash.edu},
      Daniel J. Price\(^{1}\),
      Christophe Pinte\(^{1,2}\),
      Guillaume Laibe\(^{3}\)
   }\\
   \(^{1}\)School of Physics and Astronomy, Monash University, Clayton Vic 3800,
   Australia \\
   \(^{2}\)Univ. Grenoble Alpes, CNRS, IPAG, F-38000 Grenoble, France \\
   \(^{3}\)\'{E}cole normale sup\'{e}rieure de Lyon, CRAL, UMR CNRS 5574,
   Universit\'{e} de Lyon, 46 All\'{e}e d’Italie, 69364 Lyon Cedex 07, France
 \\
}

\date{Accepted XXX. Received YYY; in original form ZZZ}

\pubyear{2020}

\begin{document}
\label{firstpage}
\pagerange{\pageref{firstpage}--\pageref{lastpage}}
\maketitle

\begin{abstract}
   We present a method for simulating the dynamics of a mixture of gas and
   multiple species of large Stokes number dust grains, typical of evolved
   protoplanetary discs and debris discs. The method improves upon earlier
   methods, in which only a single grain size could be represented, by capturing
   the differential backreaction of multiple dust species on the gas. This
   effect is greater for large dust-to-gas ratios that may be expected in the
   later stages of the protoplanetary disc life. We benchmark the method against
   analytic solutions for linear waves, drag and shocks in dust-gas mixtures,
   and radial drift in a protoplanetary disc showing that the method is robust
   and accurate.
\end{abstract}




\section{Introduction}

In order to interpret dust continuum emission, for example in ALMA observations
showing gaps and rings \citep{ALMAPartnership2015ApJ...808L...3A,
Andrews2016ApJ...820L..40A}, we must model planet-disc interactions including
multiple species. At any continuum wavelength grains of multiple sizes
contribute to the observed emission. In addition, spectral index maps can put
constraints on the size distribution of dust grains within gaps
\citep{Huang2018ApJ...852..122H}. Multi-species dust models allow us to test the
underlying disc models by producing synthetic spectral index maps to compare
with observations \citep{Casassus2015ApJ...812..126C, Pinte2016ApJ...816...25P}.

Multi-wavelength observations of protoplanetary discs show that the radial
extent of the dust disc scales inversely with wavelength, i.e.\ inversely with
dust grain size \citep{Andrews2015PASP..127..961A}. Modelling this requires dust
and gas hydrodynamics with multiple dust species---one of the grand challenges
in protoplanetary disc modelling \citep{Haworth2016PASA...33...53H}. In
particular, it is important to capture the collective backreaction of multiple
dust species on the gas \citep{Dipierro2018MNRAS.479.4187D}. For example, small
grains are well coupled to the gas and follow its inward or outward motion
\citep{Weidenschilling1977MNRAS.180...57W}. Larger grains drift inwards due to
the differential velocity between the pressure-supported gas, and, due to
conservation of angular momentum, the gas might drift outwards dragging the
small grains with it \citep{Laibe2014MNRAS.444.1940L}.

Radiative transfer calculations offer a way to validate hydrodynamical models of
protoplanetary discs by producing synthetic observations for comparison with
observations. These calculations require knowledge of the distribution of many
dust species from the hydrodynamical simulation. The ability to perform
multi-wavelength synthetic observations in single species dust-gas
hydrodynamical models is limited. One option is to stack, in post-processing,
multiple single dust species simulations \citep{Dipierro2015MNRAS.453L..73D,
Mentiplay2019MNRAS.484L.130M}. However, due to back reaction, there may be a
phase difference in the location of the concentration making this procedure
unreliable. For example, discs containing dust asymmetries and spiral arms
around central cavities can show greater concentration at different wavelengths
\citep{Casassus2015ApJ...812..126C, van-der-Marel2015ApJ...810L...7V}. Previous
work has suggested that the cause of these features is a (possibly unseen)
companion in the cavity \citep{Price2018MNRAS.477.1270P,
Poblete2019MNRAS.489.2204P,Calcino2019MNRAS.490.2579C}. Due to the potential for
a phase difference in these single-species simulations, to produce
multi-wavelength synthetic observations requires multiple dust species within
the same hydrodynamical simulation.

Several groups have developed methods for simulating the hydrodynamics of
dust-gas mixtures with multiple dust species with both grid and particle
methods, e.g. \citet{Bai2010ApJS..190..297B, Porth2014ApJS..214....4P,
Hutchison2018MNRAS.476.2186H, Benitez-Llambay2019ApJS..241...25B,
Lebreuilly2019A&A...626A..96L, Li2019ApJ...878...39L}. The methods of
\citet{Benitez-Llambay2019ApJS..241...25B} and
\citet{Lebreuilly2019A&A...626A..96L} are limited by their use of the single
fluid approach (discussed below). \citet{Bai2010ApJS..190..297B},
\citet{Porth2014ApJS..214....4P}, and \citet{Li2019ApJ...878...39L} describe
grid codes which can introduce grid-alignment issues.

Smoothed particle hydrodynamics (SPH) is a particle method for solving the
equations of hydrodynamics \citep{Monaghan1992ARA&A..30..543M,
Monaghan2005RPPh...68.1703M, Price2012JCoPh.231..759P}. The fluid, typically
gas, is discretised onto a set of particles rather than a grid. There are two
approaches to modelling of dust-gas mixtures in SPH: (i) the dust and gas are
represented by separate sets of SPH particles
\citep{Monaghan1995CoPhC..87..225M, Laibe2012MNRAS.420.2345L,
Laibe2012MNRAS.420.2365L}. In this approach, the dust and gas SPH particles
interact via a drag coupling term. In the other approach, (ii), the dust and gas
are modelled by a single set of SPH particles representing the mixture
\citep{Laibe2014MNRAS.444.1940L, Laibe2014MNRAS.440.2147L,
Laibe2014MNRAS.440.2136L, Price2015MNRAS.451..813P,
Ballabio2018MNRAS.477.2766B}. In this approach, the dust fraction is stored on
the particles and evolved in time.

Typically, in both of these methods, an explicit timestepping scheme is used.
One problem with such schemes is that the drag timescale can be many orders of
magnitude smaller than other physically interesting time scales, e.g., the
orbital time in a protoplanetary disc. This requires taking small timesteps and
is computationally inefficient. Various groups have developed implicit or
semi-implicit timestepping methods which remove the restriction of small
timesteps required by explicit schemes for drag
\citep{Monaghan1997JCoPh.138..801M, Laibe2012MNRAS.420.2365L,
Loren-Aguilar2014MNRAS.443..927L, Loren-Aguilar2015MNRAS.454.4114L,
Stoyanovskaya2018A&C....25...25S}.

In the description for both of these approaches (the dust as separate particles
method and the single SPH fluid mixture method) the methods are applicable to a
single species of dust with fixed size. \citet{Laibe2014MNRAS.444.1940L} and
\citet{Hutchison2018MNRAS.476.2186H} described a multiple species approach for
the single SPH fluid method. \citet{Hutchison2018MNRAS.476.2186H} derived the
version using the terminal velocity approximation, and tested this in the
\textsc{Phantom} SPH code \citep{Price2018PASA...35...31P}. The single fluid
approach is limited in that particles are not allowed to stream past each other
as expected for large, weakly-coupled grains \citep{Laibe2014MNRAS.440.2147L}.
The methods of \citet{Benitez-Llambay2019ApJS..241...25B,
Lebreuilly2019A&A...626A..96L, Li2019ApJ...878...39L} are based on the single
fluid approach.

In this paper we extend method (i), in which dust is represented by separate
sets of particles, from a single species to multiple species. We present the
continuum equations to solve in Section~\ref{subsec:continuum}, and the SPH
discretisation of those equations in Section~\ref{subsec:sph}. We describe
several tests of the method, as implemented in \textsc{Phantom}, in
Section~\ref{sec:tests}. We discuss some of the challenges of the method in
Section~\ref{sec:discussion}.

\section{Methods}

\subsection{Continuum equations for multiple dust species}%
\label{subsec:continuum}

We consider a mixture of gas, indexed by \(\g\), and \(N\) dust species, indexed
by \(\dd_i\). We neglect the finite size of the dust particles, and thus set the
gas volume fraction to unity. We represent each dust species as a continuous
fluid with a fixed size, \(s_i\), and intrinsic density, \(\varrho_{{\rm
m}_i}\). Then, the equations of conservation of mass for the mixture is given by
\begin{align}
   \label{eqn:conserve-gas-mass}
   \frac{\partial \rho_{\g}}{\partial t} + \nabla \cdot (\rho_{\g} \vec{v}_{\g}) &= 0, \\
   \label{eqn:conserve-dust-mass}
   \frac{\partial \rho_{\dd_i}}{\partial t} + \nabla \cdot (\rho_{\dd_i} \vec{v}_{\dd_i}) &= 0,
\end{align}
for each \(i\) in 1 to \(N\), where \(\rho_{\g}\) and \(\vec{v}_{\g}\) are the gas
density and velocity, and \(\rho_{\dd_i}\) and \(\vec{v}_{\dd_i}\) are the dust
density and velocity.

We assume the fluids are inviscid, that the dust is pressureless, and that each
dust species is homogeneous, i.e.\ has the same grain size, mass, and intrinsic
density. The equations of conservation of momentum for the mixture are given by
\begin{align}
   \rho_{\g} \frac{\dd \vec{v}_{\g}}{\dd t}
      &= - \nabla P + \rho_{\g} \vec{f} + \sum_i K_i \left(\vec{v}_{\dd_i}
         - \vec{v}_{\g}\right), \\
   \rho_{\dd_i} \frac{\dd \vec{v}_{\dd_i}}{\dd t}
      &= \rho_{\dd_i} \vec{f} - K_i \left(\vec{v}_{\dd_i} - \vec{v}_{\g}\right),
\end{align}
where \(P\) is the gas pressure, \(\vec{f}\) is any body forces acting on the
fluids, typically gravity from a star or planet (we ignore self-gravity in this
paper), and \(K_i\) is the drag coefficient between the gas and a particular
dust species, \(i\). Note that each dust fluid has one gas drag interaction
term, whereas the gas momentum equation has a sum of interactions over each dust
species. Also note that the dust has no pressure gradient force term. In
general, the drag coefficient could be a complicated expression. We assume that
the drag force is linear with respect to the differential velocity, \(\Delta
\vec{v}_i = \vec{v}_{\dd_i} - \vec{v}_{\g} \). Thus the drag coefficient is
constant in differential velocity.

The gas and dust exchange momentum via drag which leads to frictional heating.
Under the assumption that the gas and dust grains are at the same temperature,
the evolution of gas internal energy is given by
\begin{align}
   \label{eqn:conserve-energy}
   \rho_{\g} \frac{\dd u_{\g}}{\dd t} =
      - P \left(\nabla \cdot \vec{v}_{\g}\right)
      + \rho_{\g} \sum_i K_i {(\vec{v}_{\g} - \vec{v}_{\dd_i})}^2.
\end{align}
We neglect the dust fluid internal energy under the assumption that the dust and
gas are at the same temperature.

Equations~\ref{eqn:conserve-gas-mass}--\ref{eqn:conserve-energy} are \(2N + 3\)
equations describing the evolution of a mixture of gas and \(N\) dust species.
We discretise these equations with smoothed particle hydrodynamics in
Section~\ref{subsec:sph}.

\subsection{Drag timescale}

\subsubsection{Drag coefficient and stopping time}

Dust and gas interact via a drag force. This drag force has a characteristic
time scale, known as the stopping time. The stopping time relates to the drag
coefficient, which depends on quantities such as the gas temperature and
density, and on the physical characteristics of the dust grains. For a single
dust species, the stopping time, \(t_s\), is given by
\begin{align}
   \label{eqn:single-stopping-time}
   t_s = \frac{\rho_{\g} \rho_{\dd}}{K (\rho_{\g} + \rho_{\dd})},
\end{align}
where \(K\) is the drag coefficient for the single species. We assume spherical
grains of size \(s\) with a uniform material density, \(\varrho_{\rm m}\). In
the linear Epstein regime \citep{Epstein1924PhRv...23..710E} the drag
coefficient \(K\) is
\begin{align}
   \label{eqn:single-drag-coefficient}
   K = \frac{\rho_{\g} \rho_{\dd}}{\varrho_{\rm m} s} \sqrt{\frac{8}{\pi\gamma}} c_s f,
\end{align}
where \(\gamma\) is the adiabatic index of the gas. For convenience, we define
an effective material density, \(\varrho_{\mathrm{eff}}\), given by
\(\varrho_{\mathrm{eff}} = \varrho_{\rm m} \sqrt{\pi\gamma/8}\). In addition,
\(f\) is a correction for supersonic relative velocities given by
\citep{Kwok1975ApJ...198..583K}
\begin{align}
   \label{eqn:supersonic}
   f = \sqrt{1 + \frac{9\pi}{128} \frac{\Delta v^2}{c_s^2}}.
\end{align}

As discussed in \citet{Hutchison2018MNRAS.476.2186H}, a straightforward
generalisation of the stopping time for multiple dust species is not available.
Each dust species is separately coupled to the gas by the drag force. However,
the dust species are indirectly coupled to each other via backreaction, as
required by conservation of momentum. Considering the multiple dust species
case, and ignoring the supersonic correction factor, the drag coefficient,
\(K_i\), is now
\begin{align}
   \label{eqn:drag-coefficient}
   K_i = \frac{\rho_{\g} \rho_{\dd_i} c_s}{\varrho_{\mathrm{eff}} s_i}.
\end{align}
By analogy with Equation~\ref{eqn:single-stopping-time} we can define a
``stopping time'', \(t'_{s_i}\), as
\begin{align}
   \label{eqn:alternative-stopping-time}
   t'_{s_i} = \frac{\rho_{\g} \rho_{\dd_i}}{K_i (\rho_{\g} + \rho_{\dd_i})}.
\end{align}
However, there are other ``stopping times'' we can define. First, we define
\(\rho = \rho_{\g} + \sum_i \rho_{\dd_i}\) as the total density of the gas and
all dust species. Then we can define another ``stopping time'', \(t_{s_i}\), as
\begin{align}
   \label{eqn:stopping-time}
   t_{s_i} = \frac{\rho_{\g} \rho_{\dd_i}}{K_i \rho}.
\end{align}
Considering the mixture of dust and gas as a whole, we define the weighted sum
\(s_{\mathrm{eff}} = \sum_i \rho_{\dd_i} s_i / \sum_i \rho_{\dd_i}\) as an
effective grain size for the mixture. Then we can define an effective stopping
time for the dust mixture, assuming Epstein drag, as
\begin{align}
   \label{eqn:effective-stopping-time}
   T_s = \frac{\varrho_{\mathrm{eff}} s_{\mathrm{eff}}}{\rho c_s}.
\end{align}
By combining
Equations~\ref{eqn:single-stopping-time}~\&~\ref{eqn:single-drag-coefficient} we
can see the expression for the multigrain effective stopping time, \(T_s\), is
analogous to the single dust species case, \(t_s\).

\subsubsection{Stokes number}

The Stokes number, \(\mathrm{St}\), is a dimensionless stopping time defined as
the stopping time in units of a typical flow time. For protoplanetary discs, the
typical flow time is the Keplerian orbital time, \(1/\Omega_K\), so that the
Stokes number is \(\mathrm{St} \equiv t_s \Omega_K\). Note that the Stokes
number depends on the gas disc properties, via the density and temperature, the
dust disc density, the dust grain properties, i.e.\@ size and material density,
and the stellar mass and orbital distance. The disc surface density, \(\Sigma\),
and disc scale height, \(H\), are related to the density, sound speed, and
Keplerian orbital time by \(\rho = \Sigma / \sqrt{2\pi} H\) and \(\Omega_K = c_s
/ H\). Using these relations we can show that the Stokes number in the midplane
of a protoplanetary disc is given by
\begin{align}
   \mathrm{St} = \frac{\sqrt{2\pi} \varrho_{\mathrm{eff}} s}{\Sigma}.
\end{align}

Considering the multiple dust species case, we can define an effective midplane
Stokes number for the mixture using the effective stopping time
(Equation~\ref{eqn:effective-stopping-time}) as \(\mathrm{St}_{\mathrm{eff}} =
\sqrt{2\pi} \varrho_{\mathrm{eff}} s_{\mathrm{eff}} / \Sigma \). An alternative
is to define a per species midplane Stokes number in analogy with the single
grain size case:
\begin{align}
   \label{eqn:stokes}
   \mathrm{St}_i = \frac{\sqrt{2\pi} \varrho_{\mathrm{eff}} s_i}
      {\Sigma_{\g} \left(1 + \sum_i \varepsilon_i \right)},
\end{align}
where \(\Sigma_{\g}\) is the gas surface density and \(\varepsilon_i =
\rho_{\dd_i} / \rho_{\g}\) is the dust-to-gas ratio for each dust species. Again
we see the combination of grain properties, size and material density, and the
disc surface density fixing the Stokes number. Given that we assume the material
density is the same for all species, we see that the grain size of the species,
for any fixed location in the disc, gives the variation in Stokes number between
species.

The Stokes number controls the dynamics of dust grains in protoplanetary discs
\citep{Weidenschilling1977MNRAS.180...57W, Takeuchi2002ApJ...581.1344T},
affecting radial drift, vertical settling, orbital circularisation, and gap and
spiral formation. The individual Stokes number (Equation~\ref{eqn:stokes})
encapsulates the dynamics of each dust species. We can distinguish between three
regimes of dust dynamics:
\begin{enumerate}
   \item small grains, i.e.\ those with \(\mathrm{St_i} \ll 1\),
   \item intermediate-sized grains, i.e.\ those with \(\mathrm{St_i} \sim 1\),
      and
   \item large grains, i.e.\ those with \(\mathrm{St_i} \gg 1\).
\end{enumerate}
Small grains have short stopping time, i.e.\ differential velocity decays faster
than the orbital time, and are thus strongly coupled to the gas. These grains
stick to the gas, e.g.\ following the gas accretion flow. Large grains have long
stopping time, i.e.\ differential velocity decays more slowely than the orbital
time, and are thus weakly coupled to the gas. Intermediate-sized grains are
marginally coupled to the gas. These grains experience the fastest radial drift
velocities \citep{Takeuchi2002ApJ...581.1344T,Ayliffe2012MNRAS.423.1450A}. For a
typical protoplanetary disc, with surface density \(\approx 1\)~g cm\({}^{-2}\)
and intrinsic grain density \(\approx 1\)~g cm\({}^{-3}\), small grains are
\(\lesssim 10 \mu\mathrm{m}\), and large grains are \(\gtrsim 1 \mathrm{mm}\).
Note that for other physical systems the terms small and large grains have
different meaning. For example, in the ISM large grains might be any grains \(>
10 \mu\mathrm{m}\).

Several works describe the behaviour of these dust species as independently
coupled to the gas \citep{Nakagawa1986Icar...67..375N,
Dipierro2017MNRAS.469.1932D, Kanagawa2017ApJ...844..142K}.
\citet{Dipierro2018MNRAS.479.4187D} extended the analysis to consider the full
backreaction of all species onto the gas. They showed that the cumulative
backreaction from multiple dust species can strongly affect the gas flow, even
for low dust-to-gas ratio, and that, for large dust-to-gas ratio, the small
grains can drift outwards, as opposed to the typical inwards drift. This gives
motivation to the current work.

\subsection{SPH with multiple dust species}%
\label{subsec:sph}

\subsubsection{SPH density with multiple dust species}

We extend the SPH method for dust and gas mixtures with a single species, first
described in \citet{Monaghan1995CoPhC..87..225M} and improved upon by
\citet{Laibe2012MNRAS.420.2345L,Laibe2012MNRAS.420.2365L}, to multiple dust
species. We use a separate set of particles to represent the gas and each dust
species. So each phase of the mixture has its own density and smoothing length
which depend only on the neighbouring particles of its own phase. This differs
from the 1-fluid method described by \citet{Hutchison2018MNRAS.476.2186H} in
which there is a single set of SPH particles representing the mixture. The SPH
formulation of the continuity equations,
Equations~\ref{eqn:conserve-gas-mass}--\ref{eqn:conserve-dust-mass}, are given
by
\begin{align}
   \label{eqn:sph-density}
   \rho^k_a &= \sum_b m^k_b W_{ab}(h^k_b), \\
   h^k_a &= \eta \left(\frac{m^k_a}{\rho^k_a}\right)^{1/\nu},
   \label{eqn:sph-smoothing}
\end{align}
where the superscript \(k \in \{\g, \dd_1, \dots, \dd_N\}\) is an index
distinguishing between the gas and all dust species; \(\nu\) is the number of
spatial dimensions; \(m^k_a\), \(\rho^k_a\) and \(h^k_a\) are the SPH particle
mass, density, and smoothing length, respectively; \(W_{ab}\) is the SPH
smoothing kernel; and \(\eta\) is a factor of order unity which determines the
number of neighbours per particle. For an overview on the SPH method, see
\citet{Monaghan1992ARA&A..30..543M, Monaghan2005RPPh...68.1703M,
Price2012JCoPh.231..759P}. So there are \(N + 1\) sets of equations, Equations
\ref{eqn:sph-density}--\ref{eqn:sph-smoothing}, one set per phase. This is a
straightforward generalisation of previous single dust species methods.

One important feature of the above equations is that the gas and dust densities
(for each dust species) are calculated without reference to any other phase.
Given that the dust is pressureless this can lead to dust becoming trapped under
the gas resolution \citep{Laibe2012MNRAS.420.2345L}. This is in contrast to the
gas in which, due to the pressure force, particles are kept apart to prevent
this over-concentration. Dust and gas interact only via drag which requires
relative motion. If dust particles become concentrated and remain motionless
with respect to the local gas distribution there is no pressure force to stop
further collapse. This can lead to unphysical clumping of dust which can mimic
physical clumping of dust that might be expected, in, for example,
protoplanetary discs or molecular clouds. (See
\citealt{Tricco2017MNRAS.471L..52T} for a discussion of this in the context of
molecular clouds.)

\subsubsection{SPH equation of motion with multiple dust species}

The equations of motion in SPH can be derived from a Lagrangian giving exact
conservation of linear and angular momentum, and energy
\citep{Price2012JCoPh.231..759P}. Following \citet{Laibe2012MNRAS.420.2345L} and
\citet{Price2020MNRAS.495.3929P}, the equation of motion for the gas is given by
\begin{align}
   \label{eqn:sph-gas-velocity}
   \begin{split}
      \frac{\dd \vec{v}^{\g}_a}{\dd t} = &- \sum_{b \in g} m^{\g}_b \left[
         \frac{P_a + q_{ab}^a}{\Omega^{\g}_a {\left(\rho^{\g}_a\right)}^2} \nabla_a W_{ab}(h^{\g}_a) +
         \frac{P_b + q_{ab}^b}{\Omega^{\g}_b {\left(\rho^{\g}_b\right)}^2} \nabla_a W_{ab}(h^{\g}_b)
      \right] \\
      &+ \nu \sum_{i=1}^N \sum_{b \in d_i} m^{\g}_b \frac{K_{ab}}{\rho^{\g}_a \rho^{\dd_i}_b}
         (\vec{v}^{*}_{ab} \cdot \hat{\vec{r}}_{ab}) \hat{\vec{r}}_{ab} D_{ab}(h_{\max}),
   \end{split}
\end{align}
where \(K_{ab}\), \(D_{ab}\), \(h_{\max}\) and \(\vec{v}^{*}_{ab}\) are defined
below. \(P_a\) refers to the pressure on particle \(a\), \(\Omega^{\g}_a\) is a
term related to the variable smoothing length given by
\begin{align}
   \Omega^{\g}_a = 1 - \frac{\partial h_a}{\partial \rho_a}
      \sum_b m_b \frac{\partial W_{ab}(h_a)}{\partial h_a}.
\end{align}
\(q_{ab}^a\) and \(q_{ab}^b\) are terms relating to artificial viscosity given
by
\begin{align}
   q_{ab}^a =
   \begin{cases}
      -\frac{1}{2} \rho_a v_{{\rm sig}, a} \vec{v}_{ab} \cdot \hat{\vec{r}}_{ab},
         &\vec{v}_{ab} \cdot \hat{\vec{r}}_{ab} < 0,\\
      0, &\mathrm{otherwise},
   \end{cases}
\end{align}
where \(\hat{\vec{r}}_{ab} = (\vec{r}_a - \vec{r}_b)/|\vec{r}_a - \vec{r}_b|\)
is the normalised separation vector, \(\vec{v}_{ab} = \vec{v}_a -
\vec{v}_b\), and \(v_{{\rm sig}, a}\) is the maximum signal speed given by
\begin{align}
   v_{{\rm sig}, a} = \alpha_a^{\rm AV} c_{s,a}
      + \beta^{\rm AV} | \vec{v}_{ab} \cdot \hat{\vec{r}}_{ab} |,
\end{align}
where \(\alpha_a^{\rm AV}\) and \(\beta^{\rm AV}\) are parameters controlling
the artificial viscosity strength. The first term in
Equation~\ref{eqn:sph-gas-velocity} represents the usual SPH discretisation of
the pressure gradient force on a particle, labelled by \(a\). The sum, denoted
by \(\sum_{b \in g}\), is a sum over gas neighbours, labelled by \(b\). The
second term is discussed in Section~\ref{sec:drag-kernel}.

The equation of motion for each dust species, \(\dd_i\), is given by
\begin{align}
   \label{eqn:sph-dust-velocity}
   \frac{\dd \vec{v}^{\dd_i}_a}{\dd t} =
      - \nu \sum_{b \in g} m^{\dd_i}_b \frac{K_{ab}}{\rho^{\g}_b \rho^{\dd_i}_a}
      (\vec{v}^{*}_{ab} \cdot \hat{\vec{r}}_{ab}) \hat{\vec{r}}_{ab} D_{ab}(h_{\max}),
\end{align}
where \(K_{ab}\) is the drag coefficient between particles \(a\) and \(b\),
\(D_{ab}\) is the drag kernel (Section~\ref{sec:drag-kernel}), \(h_{\max} =
\max(h_a, h_b)\) is the maximum smoothing length between any pair of particles,
and \(\vec{v}^{*}_{ab}\) refers to a higher-order reconstructed velocity for the
particle pair \(a\) and \(b\) \citep{Price2020MNRAS.495.3929P}. This avoids the
``overdamping'' problem described by \citet{Laibe2012MNRAS.420.2345L}.

\subsubsection{Drag kernel and drag coefficient}
\label{sec:drag-kernel}

The second term in Equation~\ref{eqn:sph-gas-velocity} represents the cumulative
drag force from each dust species on particle \(a\) (with \(i \in \{1, \ldots,
N\}\) representing each species). For each species, the sum, denoted by
\(\sum_{b \in d_i}\), is a sum over dust neighbours, labelled by \(b\), where
here the \(a\) index refers to the gas particle and the \(b\) index refers to
its dust neighbours. Similarly, the sum on the right hand side of
Equation~\ref{eqn:sph-dust-velocity} represents the drag force from the gas on
the particular dust species, where \(a\) refers to a dust (SPH) particle and
\(b\) refers to its gas neighbours.

Equations~\ref{eqn:sph-gas-velocity} and~\ref{eqn:sph-dust-velocity} are a
straightforward generalisation of \citet{Laibe2012MNRAS.420.2345L} to multiple
dust species. For Epstein drag, the drag coefficient \(K_{ab}\) is given by
\begin{align}
   K_{ab} = \frac{\rho^{\g}_a \rho^{\dd_i}_b}{\varrho_{\mathrm{eff}} s_i} c_{s,a} f_{ab},
\end{align}
where \(c_{s,a}\) is the sound speed on the gas particle and \(f_{ab}\)
supersonic correction factor (Equation~\ref{eqn:supersonic}) between the two
particles.

The typical bell-shaped kernel used in SPH to estimate density, denoted by
\(W_{ab}\), is not appropriate for the drag force summation. Rather, we use a
separate drag kernel, \(D_{ab}\), with a ``double-hump''. This was found to be
10x more accurate \citep{Laibe2012MNRAS.420.2345L}.

\subsection{Time stepping}

\subsubsection{Time step constraint}

We use an explicit leapfrog time stepping scheme as discussed in
\citet{Price2018PASA...35...31P}. Given that we use an explicit scheme for the
drag term there is an additional stability constraint on the timestep \(\Delta
t\). We derive this constraint for the forward Euler scheme. (Even though we do
not use this scheme, it provides a guide to the nature of the constraint.)
Following \citet{Laibe2012MNRAS.420.2345L, Laibe2014MNRAS.444.1940L}, we start
with the discretisation of the drag-only velocity equations for dust and gas
with the forward Euler method, we have
\begin{align}
   \frac{v_{\g}^{n+1} - v_{\g}^n}{\Delta t} &= - \sum_i \frac{K_i}{\rho_{\g}} \left(v_{\g}^n - v_{\dd_i}^n\right), \\
   \frac{v_{\dd_i}^{n+1} - v_{\dd_i}^n}{\Delta t} &= \frac{K_i}{\rho_{\dd_i}} \left(v_{\g}^n - v_{\dd_i}^n\right).
\end{align}
Then we perform a von Neumann stability analysis, where we expand the solution
in plane waves, \(v_j^m = V_j^m e^{i k x}\), where \(j\) refers to the species,
\(k\) is the wavenumber, and \(m\) refers to the time step. This expansion leads
to the following matrix equation:
\begin{align}
   \begin{pmatrix}
      V_{\g} \\ V_{d_1} \\ \vdots \\ V_{d_N} \\
   \end{pmatrix}^{n+1} =
   \begin{pmatrix}
      1 - \frac{\Delta t \sum K_i}{\rho_{\g}} & \frac{\Delta t K_1}{\rho_{\g}} & \cdots & \frac{\Delta t K_N}{\rho_{\g}} \\
      \frac{\Delta t K_1}{\rho_{d_1}} & 1 - \frac{\Delta t K_1}{\rho_{d_1}} & \cdots & 0 \\
      \vdots & \vdots & \vdots & \vdots \\
      \frac{\Delta t K_N}{\rho_{d_N}} & 0 & \cdots & 1 - \frac{\Delta t K_N}{\rho_{d_N}} \\
   \end{pmatrix}
   \begin{pmatrix}
      V_{\g} \\ V_{d_1} \\ \vdots \\ V_{d_N} \\
   \end{pmatrix}^{n}.
\end{align}
To find the eigenvalues of the matrix \(M\) above we need to find the
characteristic polynomial, \(\det(M - \lambda I)\). We can find the determinant
of \(M\) using Schur's determinant identity \(\det(M) = \det(A - B D^{-1} C)
\det(D)\), where \(A\) is the 1x1 matrix consisting of the upper left value,
\(B\) is the \(N\)x1 matrix of the top row except the first value, \(C\) is the
1x\(N\) matrix of the left column except the first value, and \(D\) is the
remaining \(N\)x\(N\) diagonal matrix. Applying this identity to the matrix \(M
- \lambda I\) gives
\begin{align}
   \left(1 - \frac{\Delta t \sum K_i}{\rho_{\g}} - f - \lambda\right)
   \times \prod_i \left(1 - \frac{\Delta t K_i}{\rho_{\dd_i}} - \lambda\right),
\end{align}
where \(f = \sum_i \Delta t^2 K_i^2 \left[\rho_{\g} \rho_{\dd_i} \left( 1 -
\frac{\Delta t K_i}{\rho_{\dd_i}} - \lambda \right) \right]^{-1}\). We neglect
\(f\) as it is second order in \(\Delta t\). To find the eigenvalues we equate
the characteristic polynomial with zero and solve to find
\begin{align}
   \lambda_{\g} = 1 - \frac{\Delta t \sum_i K_i}{\rho_{\g}} \quad \mathrm{and} \quad
   \lambda_{\dd_i} = 1 - \frac{\Delta t K_i}{\rho_{\dd_i}}.
\end{align}
For stability, we require \(|\lambda_k| < 1\). Using this condition with
Equations~\ref{eqn:drag-coefficient}~\&~\ref{eqn:stopping-time} gives the
following timestep criteria:
\begin{align}
   \Delta t < \frac{2}{\sum_i \epsilon_i / t_{s_i}} \quad {\rm and} \quad
   \Delta t < \frac{2 t_{s_i}}{1 - \epsilon} \quad \forall \, i,
\end{align}
where \(\epsilon = \sum_i \epsilon_i\) and \(\epsilon_i = \rho_{\dd_i} / \rho\).
We have \(N + 1\) inequalities. We take the minimum over dust stopping times in
the inequality on the right to find the most restrictive timestep criterion
involving only dust quantities:
\begin{align}
   \label{eqn:timestep-dust}
   \Delta t < \frac{2}{1 - \epsilon} \min_i t_{s_i}.
\end{align}
\citet{Laibe2014MNRAS.444.1940L} derived more general bounds on the eigenvalues
of the drag matrix which imply a timestep criterion
\begin{align}
   \Delta t < \left( \max_i \left(\frac{1}{\epsilon_i t'_{s_i}}\right)
      + \frac{1}{1 - \epsilon} \sum_i t'_{s_i} \right)^{-1}.
\end{align}
Assuming strong drag, this can be approximated by
\begin{align}
   \Delta t < \left( \frac{1}{\min_i t_{s_i}}
      + \frac{1}{(1 - \epsilon) t'_{s_i}} \right)^{-1}.
\end{align}
We take the minimum over dust stopping times in the first term of the inequality
to find the most restrictive timestep criterion given by
\begin{align}
   \label{eqn:timestep-gas}
   \Delta t < \frac{2}{N} \frac{\max_i t_{s_i}}{\min_i \epsilon_i}.
\end{align}
This is a sufficient but not necessary condition. Given that \(N\) is of order
tens and, at worst, \(\min_i \epsilon_i \sim 1\), we can say that
Equation~\ref{eqn:timestep-dust} is more restrictive than
Equation~\ref{eqn:timestep-gas}, and thus provides the time step constraint for
drag. Note that (i) the stability constraint depends on the shortest stopping
time, and (ii) for large dust-to-gas ratios, i.e. \(\epsilon \rightarrow 1\),
the stability constraint becomes unimportant.

This timestep constraint is not the same as the one derived in
\citet{Laibe2012MNRAS.420.2345L} for a single dust species. In that case, the
time step constraint is the stopping time itself, and does not depend on the
dust fraction. Note that we have shown that Equation~\ref{eqn:timestep-dust} is
the stability constraint for the forward Euler discretisation of the drag terms
to first order in \(\Delta t\). We have not derived the stability constraint for
the leapfrog scheme we use in our numerical tests below, but we expect the
stability constraint to remain the same.

\subsubsection{Time step constraint for SPH}

We can rewrite the time step constraint in Equation~\ref{eqn:timestep-dust}
using the alternative stopping time, \(t'_{s_i}\) (defined in
Equation~\ref{eqn:alternative-stopping-time}). First we note that \(t_{s_i} / (1
- \epsilon) = \rho_{d_i} / K_i\) and \(t'_{s_i} (1 + \varepsilon_i) = \rho_{d_i}
/ K_i\). Then, given that \(1 - \epsilon\) does not depend upon the index \(i\),
we can bring it under the min operator to give
\begin{align}
   \Delta t < 2 \min_i \{t'_{s_i} (1 + \varepsilon_i)\}.
\end{align}
This is useful because in evolving the momentum equations we use the drag
coefficient on a dust (or gas) particle over its gas (or dust) neighbours,
without reference to the other dust species. Given that \(\varepsilon_i > 0\),
for simplicity, we take the more restrictive constraint of setting
\(\varepsilon_i = 0\), i.e.
\begin{align}
   \label{eqn:sph-timestep-constraint}
   \Delta t < 2 \min_i t'_{s_i}.
\end{align}
Even though using a factor in Equation~\ref{eqn:sph-timestep-constraint} of 2
provides a numerically stable timestepping scheme (in the von Neumann sense),
numerical experimentation showed that any value greater than 1.0 led to
inaccurate results. We use \(C_{\rm drag}\) to refer to this factor.

For a dust particle, we take the minimum over gas neighbours (labelled by a) of
the individual species constraint (\(\Delta t < C_{\rm drag} t_{s_i}\)):
\begin{align}
   \Delta t < C_{\rm drag} \min_{a}
   \left\{ \frac{\rho^{\g}_a \rho^{\dd_i}_b}{K_{ab} (\rho^{\g}_a + \rho^{\dd_i}_b)} \right\}.
\end{align}
Whereas, for a gas particle, we take the minimum of
Equation~\ref{eqn:sph-timestep-constraint}, rewritten in terms of the drag
coefficient and densities, over the neighbouring dust particles (labelled by b)
to find the SPH time step constraint for a particle:
\begin{align}
   \Delta t < C_{\rm drag} \min_i \min_{b}
   \left\{ \frac{\rho^{\g}_a \rho^{\dd_i}_b}{K_{ab} (\rho^{\g}_a + \rho^{\dd_i}_b)} \right\}.
\end{align}

\section{Numerical tests}%
\label{sec:tests}

We implemented the numerical method described in Section~\ref{subsec:sph} in the
smoothed particle hydrodynamics code \textsc{Phantom}\footnote{See the git
commit labelled by the hash
\href{https://github.com/danieljprice/phantom/commit/64dbd2b124ca74051eed920d6cad0a2e83157478}{\texttt{64dbd2b1}}
in the \textsc{Phantom} source code repository for the implemented changes.}. We
then performed several tests to validate the method against known analytical
solutions. The dusty box test (Section~\ref{subsec:box}) validates the drag
force coupling in the absence of spatial gradients. The dusty wave test
(Section~\ref{subsec:wave}) validates the method with spatial gradients in a
linear regime. The dusty shock test (Section~\ref{subsec:shock}) validates the
method in a challenging non-linear regime. The radial drift test
(Section~\ref{subsec:radialdrift}) validates the method is the context of a
global 3-dimensional protoplanetary disc model.

The tests all have analytical solutions in one spatial dimension. We performed
each of the tests in three dimensions, and then reduced the data to one
dimension for comparison with the analytical solutions. In each of the tests we
used, unless stated otherwise: a globally isothermal equation of state; the
quintic kernel with 113 mean particle neighbours (\(\eta = 1.0\) in
Equation~\ref{eqn:sph-density}); global timestepping, i.e., all particles have
the same time step; a Courant factor of 0.3; a time step constraint on the
acceleration \( C_{\rm force} \) of 0.25, where \(\Delta t_a < C_{\rm force}
\sqrt{h_a / |\vec{a}_a|} \); and a time step constraint on the drag \(C_{\rm
drag}\) of 0.9.

\subsection{Dusty box}%
\label{subsec:box}

\begin{figure*}
   \begin{center}
      \includegraphics[width=\textwidth]{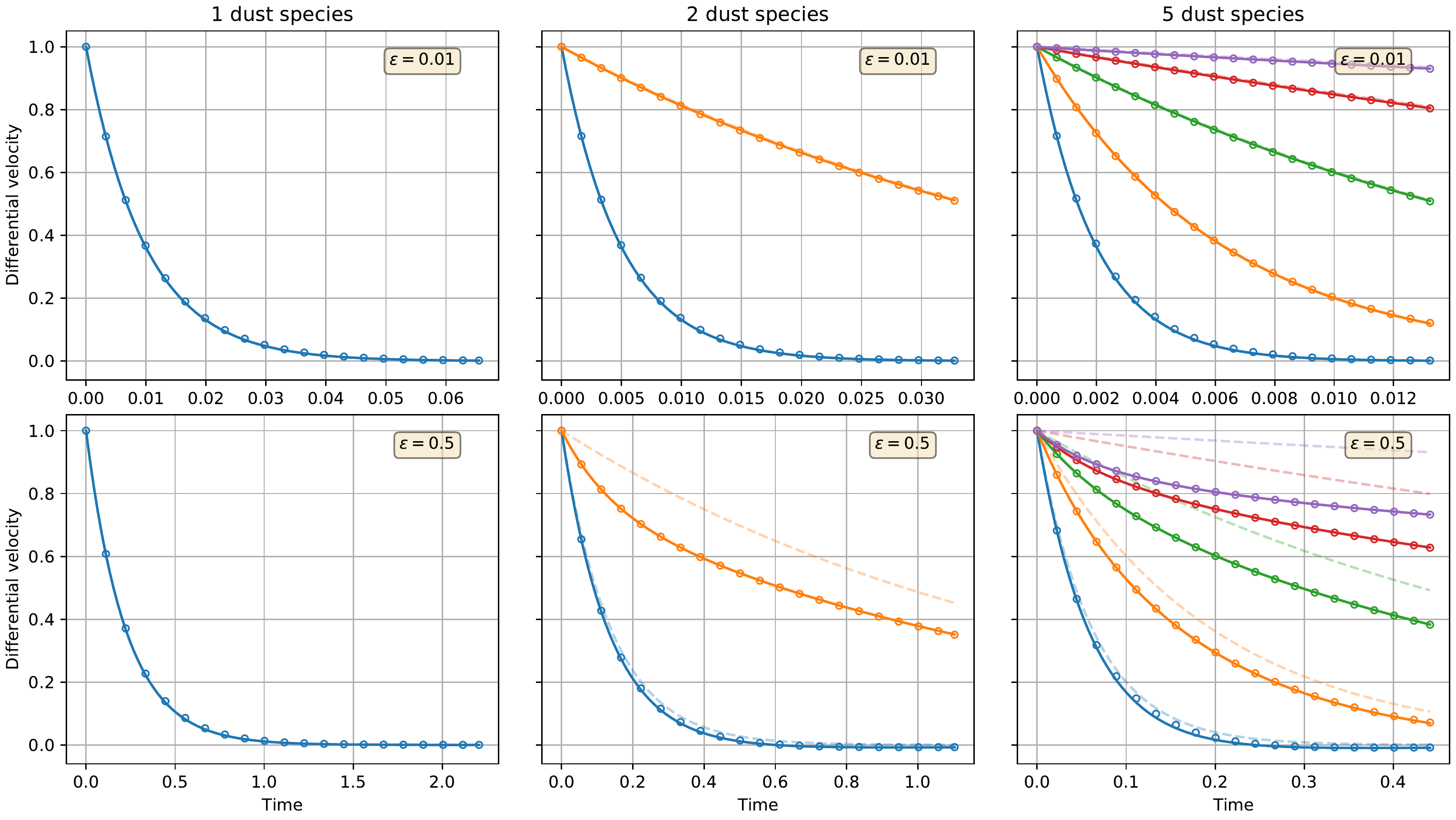}
      \caption{Dusty box numerical test showing the differential velocity
         between the dust and gas. The total dust-to-gas ratio is 0.01 (top row)
         and 0.5 (bottom row). From left to right: the number of dust species is
         1, 2, 5. The open circles represent the results from the
         \textsc{Phantom} numerical solution. The solid and dashed lines
         represent the analytical solution with and without taking back reaction
         into account, respectively. In the top row, the solid and dash lines
         lie on top of each other. In the bottom row, the numerical solution
         matches the analytical solution including backreaction. Each colour
         represents the differential velocity of a dust species increasing in
         size from bottom to top. For the specific grain sizes, see
         Table~\ref{tab:box}. Time is dimensionless, scaled by the shortest
         stopping time.%
         \label{fig:dustybox}}
   \end{center}
\end{figure*}

\begin{figure}
   \begin{center}
      \includegraphics[width=\columnwidth]{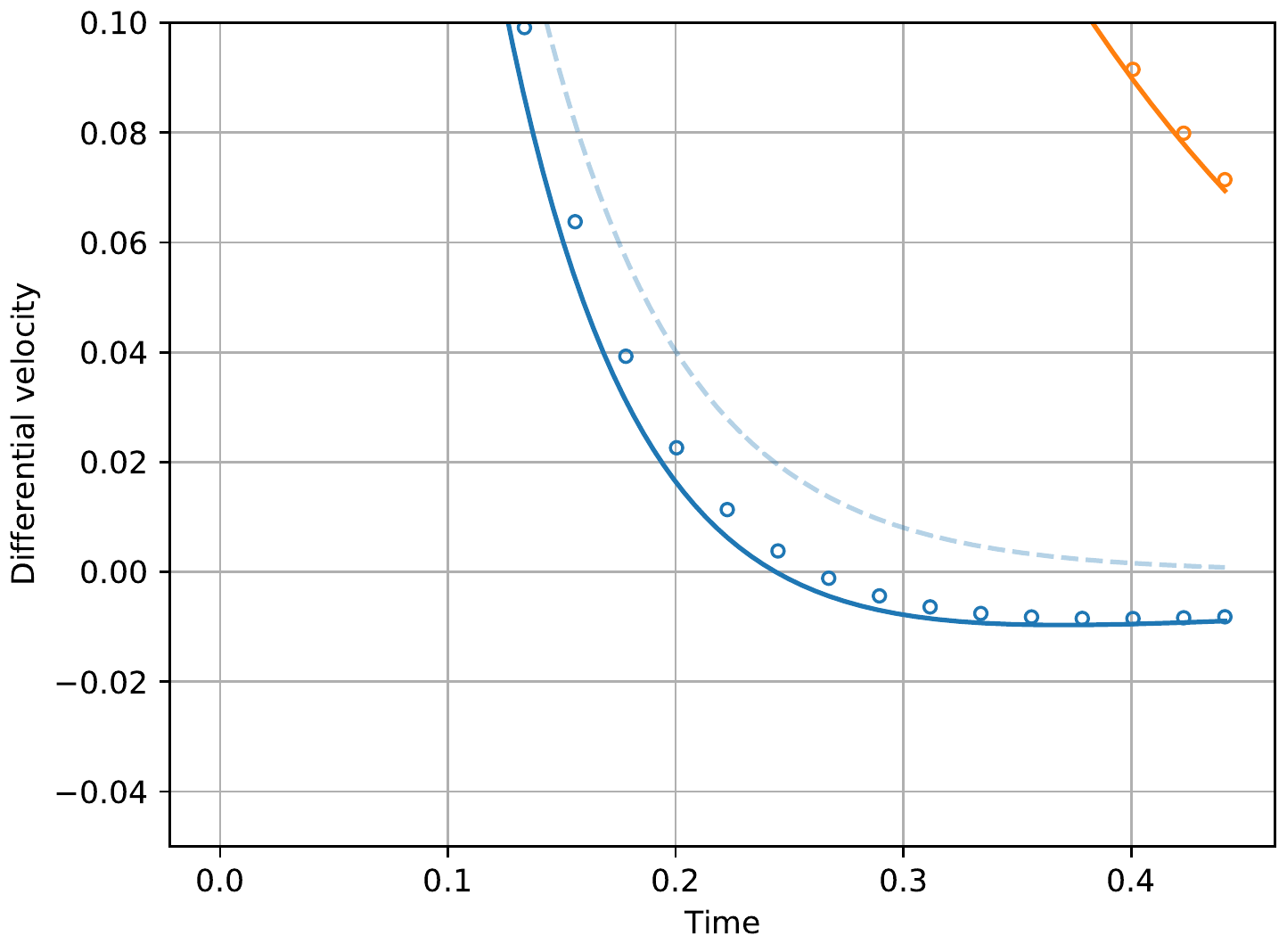}
      \caption{Zoomed in version of bottom right panel in
         Figure~\ref{fig:dustybox}. The smallest dust grains (in blue) show a
         negative differential velocity.%
         \label{fig:dustybox-zoomed}}
   \end{center}
\end{figure}

\begin{figure}
   \begin{center}
      \includegraphics[width=\columnwidth]{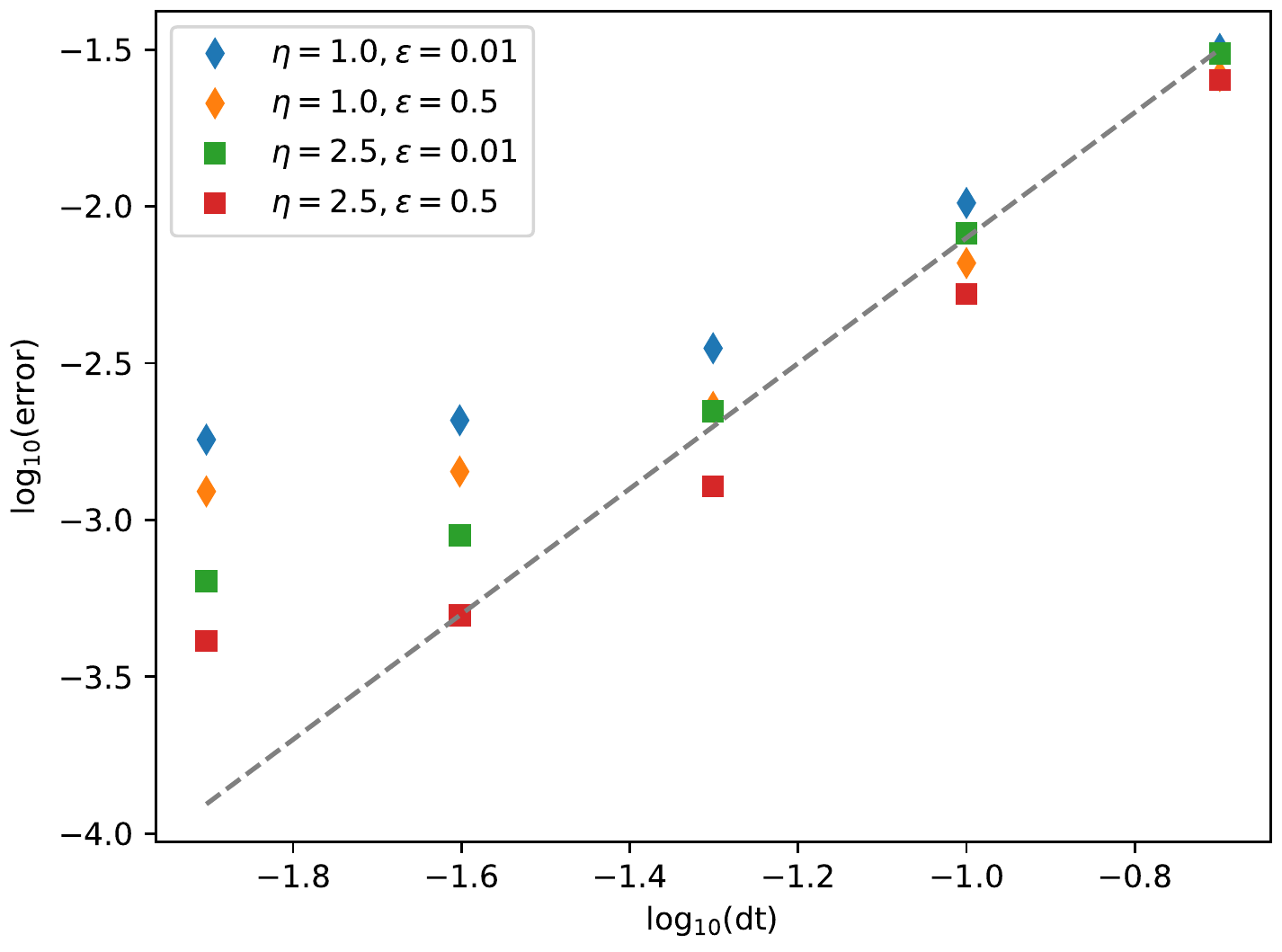}
      \caption{The L2-norm error with varying timestep showing the second order
         convergence of the method. Each marker represents a single simulation
         with two dust species with a dust-to-gas ratio of either 0.01 or 0.5.
         The gray dashed line represents a slope of 2. For large timesteps the
         error matches the line of slope 2 (indicating second-order
         convergence). For smaller timesteps the error is dominated by kernel
         bias. The diamond and square markers represent numerical solutions with
         \(\eta=1.0\) and \(\eta=2.5\) respectively.%
         \label{fig:dustybox-accuracy}}
   \end{center}
\end{figure}

We performed the multigrain version of the dusty box test described in
\citet{Laibe2011MNRAS.418.1491L}. We set up a periodic box of uniform density
gas and dust with an initial differential velocity between the gas and each dust
species. In this test, the equation of motion simplifies to
\begin{align}
   \frac{\partial \Delta \vec{V}}{\partial t} = - \Omega \Delta \vec{V},
\end{align}
where \(\Delta \vec{V}\) is the differential velocity vector in the direction of
motion, i.e. \(\vec{v}_{\dd_i} - \vec{v}_{\g}\) for each \(i\) projected along
the direction of motion, and \(\Omega\) is the drag matrix (Equation~65 of
\citealt{Laibe2014MNRAS.444.1940L}) given by,
\begin{align}
   \Omega_{ij} =
   \begin{cases}
      \frac{1}{t''_{s_i}} \frac{1}{(1 - \varepsilon)}, &i \neq j,\\
      \frac{1}{t''_{s_i}} \left( \frac{1}{\varepsilon_i} +
         \frac{1}{1 - \varepsilon} \right), &i = j,
   \end{cases}
\end{align}
where \( \varepsilon = \sum_i \varepsilon_i \), and \( t''_{s_i} = \left( \sum_i
\rho_i \right) / K_i \).

\begin{table}
   \centering
   \begin{tabular}{ccc}
      \hline
      \hline
      Species & Grain size [cm] \\
      \hline
      \hline
      One dust species \\
      1 & 0.01 \\
      \hline
      Two dust species \\
      1 & 0.01 \\
      2 & 0.1 \\
      \hline
      Five dust species \\
      1 & 0.01 \\
      2 & 0.0316 \\
      3 & 0.1 \\
      4 & 0.316 \\
      5 & 1.0 \\
      \hline
      \hline
   \end{tabular}
   \caption{Grain sizes for dusty box test.}%
   \label{tab:box}
\end{table}

This problem tests how well the numerical scheme captures the exchange of
momentum between gas and each dust species via the drag force. All tests were in
the linear Epstein drag regime. We performed six tests in two sets of three. The
first set had a total dust-to-gas ratio of 0.01, and the second set 0.5. The
tests within each set had 1, 2, and 5 dust species, respectively, with grain
sizes given in Table~\ref{tab:box}. Each test had equal mass in each grain size
bin. The gas is initially motionless, and each dust species has uniform velocity
in the positive x-direction. We turned off the SPH viscosity, i.e.\@ set
\(\alpha_{\rm AV}\) to zero in \textsc{Phantom}. We set \(C_{\rm drag} = 0.2\)
for computational reasons (to allow \textsc{Phantom} to output data more
frequently while still restricting the timestep via the drag force).

For each test problem, we set up each of the gas and dust fluids on a
close-packed lattice, using dense-sphere packing, such that there were 648
particles per species (8 particles in the direction of motion). We set the gas
density to \(10^{-13}\)~g~cm\({}^{-3}\), with the dust-to-gas ratio varying per
test problem. We set the dust grain material density for all species to
1.0~g~cm\({}^{-3}\).

Figure~\ref{fig:dustybox} shows the time evolution of the mean velocity
differential between the gas and each dust species, compared with analytical
solutions. The time is a dimensionless time scaled by the stopping time,
\(t_{s_i}\), of the smallest grain size, which is 9.677, 19.354, and
48.385~years for the low dust-to-gas ratio (0.01) calculations (top row), and
0.287, 0.575, and 1.437~years for the high dust-to-gas ratio (0.5) calculations
(bottom row). Given that there are no spatial gradients in the problem, all
particles follow the mean velocity. The dashed lines represent the analytical
solution without backreaction from the dust on the gas. The solid lines
represent the analytical solution including backreaction.

For low dust-to-gas ratio (0.01) both analytical solutions give the same decay
of differential velocity, with which the \textsc{Phantom} simulation agrees. For
a larger dust-to-gas ratio (0.5) the analytical solutions for multiple species
differs from the single-species solution, and the \textsc{Phantom} simulation
data follows the backreaction-inclusive solution. In all cases, the numerical
solution matches the analytical solution with relative error less than 0.01.

Figure~\ref{fig:dustybox-zoomed} shows a zoomed version of the bottom right
panel in Figure~\ref{fig:dustybox}. For large dust-to-gas ratio we see that the
smallest grains (0.01 cm) rapidly slow and the differential velocity reverses
sign. That is the small grains slow to the gas velocity; then, as the larger
grains speed up the gas via drag, the gas drags the small grains along with it.
This shows that the behaviour of multiple dust species for large dust-to-gas
ratios requires taking back reaction into consideration to capture the physics
of dust drag accurately \citep{Gonzalez2017MNRAS.467.1984G,
Dipierro2018MNRAS.479.4187D}.

Figure~\ref{fig:dustybox-accuracy} shows that the method is second-order
accurate with respect to the timestep. Each marker represents the L2-norm error,
of a numerical solution for the two dust species case with a dust-to-gas ratio
of either 0.01 or 0.5, with varying timestep controlled by \(C_{\rm drag}\). The
error is given by
\begin{align}
   \mathrm{error}(\Delta t) = \sqrt{ \sum_{i,j}
      \left(V_{i,j}^{\Delta t} - v_{j}(t_i) \right)^2 },
\end{align}
where the indices i and j represent time and dust species, respectively,
\(V_{i,j}^{\Delta t}\) is the numerical solution for a particular \(\Delta t\)
(or equivalently \(C_{\rm drag}\)), and \(v_{j}(t_i)\) represents the exact
solution. For large timesteps the error matches the gray, dashed line of slope 2
(indicating second-order convergence in the log-log plot). For smaller timesteps
the error is dominated by kernel bias. This is demonstrated by the fact that in
the case with fewer neighbours (\(\eta = 1.0\)), indicated by diamond markers,
the error plateaus at a larger magnitude than for the case with a greater number
of neighbours (\(\eta = 2.5\)), indicated by square markers.

\subsection{Dusty wave}%
\label{subsec:wave}

\begin{figure*}
   \begin{center}
      \includegraphics[width=0.66\textwidth]{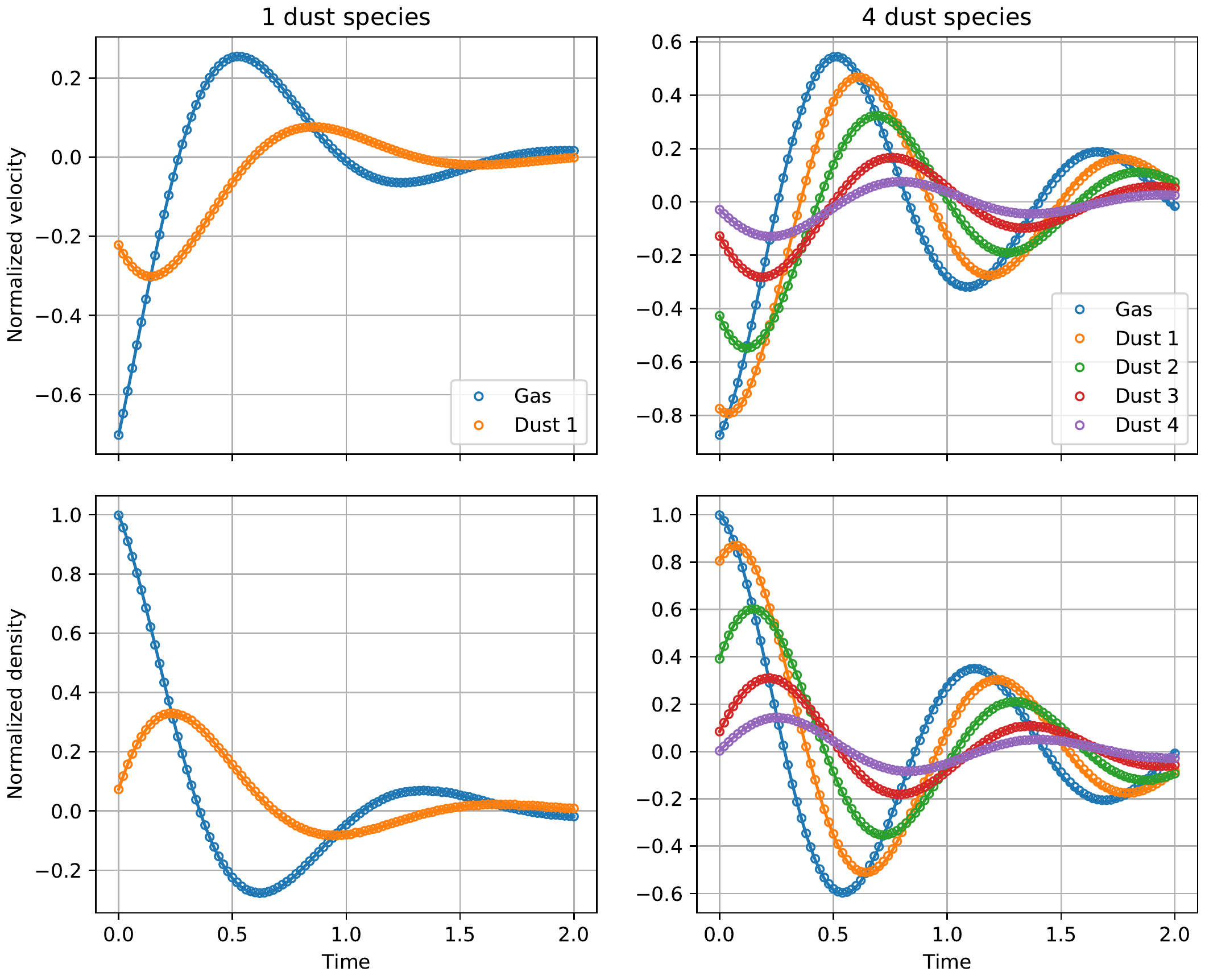}
      \caption{Dusty wave numerical test showing the normalised velocity (top)
         and density (bottom) perturbation at \(x = 0\), for gas with one dust
         species (left) and gas with four dust species (right). The open circles
         represent the \textsc{Phantom} simulation, and the solid line
         represents the analytical solution from
         \citet{Benitez-Llambay2019ApJS..241...25B}. The gas is blue, and other
         colours represent different dust species.%
         \label{fig:dustywave}}
   \end{center}
\end{figure*}

We performed the multigrain version of the dusty wave test described in
\citet{Laibe2011MNRAS.418.1491L, Laibe2014MNRAS.444.1940L}. This is a test of
the dust-gas drag coupling in the context of a damped sound wave. The dust is a
pressureless fluid and cannot support sound waves. However, the gas can support
sounds waves and drags the dust. This leads to damping of the wave. This tests
the numerical method in a problem including spatial gradients in a linear
regime.

Considering small perturbations around an equilibrium state, \(\rho_j = \rho_j^0
+ \delta \rho_j\), and \(v_j = \delta v_j\), where the index \(j\) is \(\g\) for
the gas species, and \(\dd_i\) for each of the dust species, the linearised
equations of motion for this system are
\begin{align}
   \frac{\partial \delta \rho_{\g}}{\partial t}
      + \rho_{\g}^0 \frac{\partial \delta v_{\g}}{\partial x} &= 0, \\
   \frac{\partial \delta \rho_{\dd_i}}{\partial t}
      + \rho_{\dd_i}^0 \frac{\partial \delta v_{\dd_i}}{\partial x} &= 0, \\
   \rho_{\g}^0 \frac{\partial \delta v_{\g}}{\partial t}
      &= \sum_i K_i \left(\delta v_{\dd_i} - \delta v_{\g} \right)
         + c_s^2 \frac{\partial \delta \rho_{\g}}{\partial x}, \\
   \rho_{\dd_i}^0 \frac{\partial \delta v_{\dd_i}}{\partial t}
      &= - K_i \left(\delta v_{\dd_i} - \delta v_{\g}\right).
\end{align}
For the particular setup we followed \citet{Benitez-Llambay2019ApJS..241...25B}.
By assuming solutions to the linearised equations of the form \(\delta f =
\delta \hat{f} e^{ikx - i\omega t}\) they derive solutions as a dispersion
relation (their Equation~45) and associated set of eigenfunctions (their
Equations~46--48). Following \citet{Benitez-Llambay2019ApJS..241...25B}, we set
the initial condition to constant density and zero velocity plus a perturbation
of the form
\begin{align}
   \delta f = A \left[\mathrm{Re} \left(\delta \hat{f} \right) \cos(kx)
      - \mathrm{Im} \left(\delta \hat{f} \right) \sin(kx) \right].
\end{align}
We set the sound speed \(c_s = 1\), and the wave amplitude \(A\) to \(10^{-4}
c_s\) and \(10^{-4} \rho_{\g}^0\) for the velocity and density perturbations,
respectively.

\begin{table}
   \centering
   \begin{tabular}{ccc}
      \hline
      \hline
      Species & Density & Stopping time \\
      \hline
      \hline
      One dust species \\
      g & 1.0 & - \\
      1 & 2.24 & 0.4 \\
      \hline
      Four dust species \\
      g & 1.0 & - \\
      1 & 0.1 & 0.1 \\
      2 & 0.2333 & 0.2154 \\
      3 & 0.3667 & 0.4642 \\
      4 & 0.5 & 1.0 \\
      \hline
      \hline
   \end{tabular}
   \caption{Mean density and stopping times for dusty wave test.}%
   \label{tab:wave}
\end{table}

We performed two tests: (1) with gas and a single dust species, and (2) with gas
and four dust species. We set the background density, drag coefficients, and
initial perturbations from Table~2 in
\citet{Benitez-Llambay2019ApJS..241...25B}. We set up a periodic box of unit
length, with 8192 particles for each species (128 particles in the wave
direction). We use constant drag, \(K_i\), for each dust species, where \(K_i =
\rho_{\dd_i} / t_{s_i}\), and \(\rho_{\dd_i}\) and \(t_{s_i}\) from
Table~\ref{tab:wave} (following \citealt{Benitez-Llambay2019ApJS..241...25B}).
We turned off the SPH viscosity, i.e. set \(\alpha_{\rm AV}\) to zero in
\textsc{Phantom}.

Figure~\ref{fig:dustywave} shows the time evolution of the normalised velocity
and density perturbations at a particular location within the domain (\(x=0\)).
The normalised velocity \( v_N \) and density \( \rho_N \) are defined by
\begin{align}
   \rho_N &= \frac{\overline{\rho} - \overline{\rho}_0}{A \overline{\rho}_0}, \\
   v_N &= \frac{\overline{v}}{A c_s},
\end{align}
where the zero subscript represents the initial value. The solid lines represent
the analytical solution from \citet{Benitez-Llambay2019ApJS..241...25B} and the
open circles represent the numerical solution from \textsc{Phantom}. We see that
the numerical solution accurately reproduces the analytical solution, i.e.\@ the
relative error is everywhere less than 0.01. In both cases we see the wave
damping effect, and within a few wave periods the wave dissipates.

\subsection{Dusty shock}%
\label{subsec:shock}

\begin{figure}
   \begin{center}
      \includegraphics[width=\columnwidth]{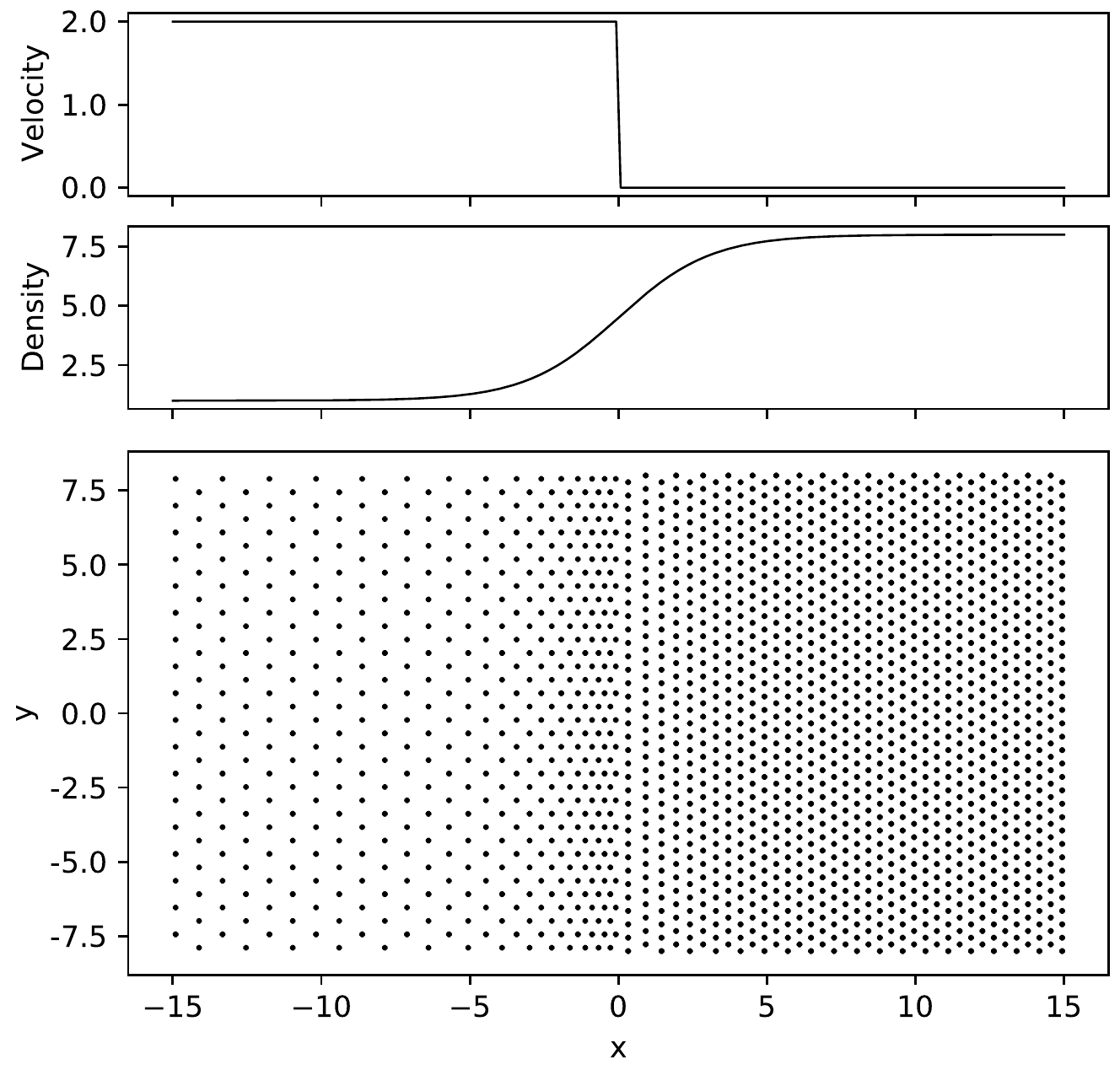}
      \caption{Dusty shock initial conditions. We used a step function for the
         velocity profile (top), and a smoothed density profile (middle). The
         particles for each species are placed on two close packed lattices, one
         on either side of the shock (bottom). The density is lower in the
         pre-shock region, on the left, than on the right. The particle
         positions in the shock region are adjusted to smooth the shock over a
         few particle spacings.%
         \label{fig:dustyshock_initial}}
   \end{center}
\end{figure}

\begin{figure*}
   \begin{center}
      \includegraphics[width=0.66\textwidth]{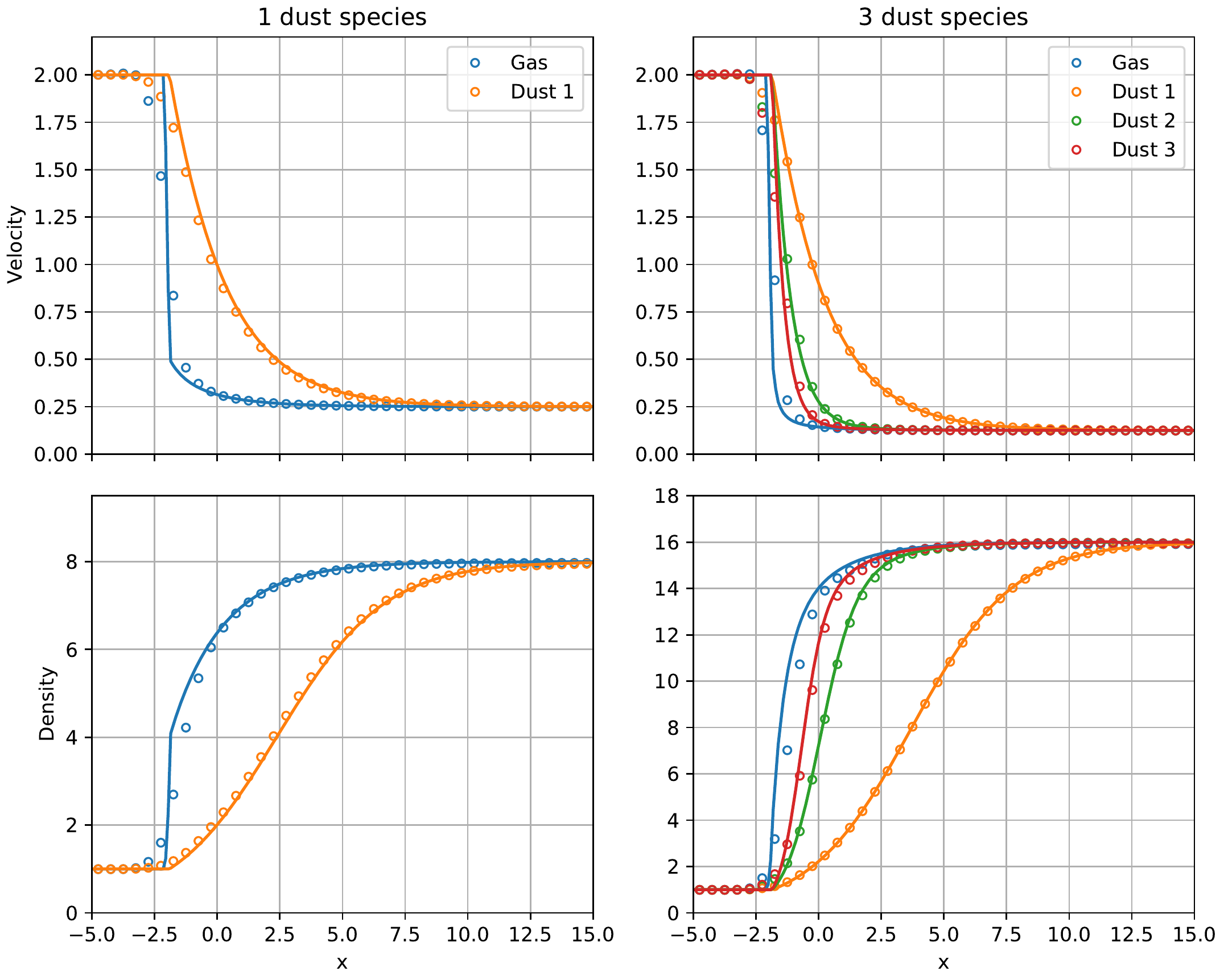}
      \caption{Dusty shock numerical test showing the velocity (top) and the
         density (bottom) at time \(t=300\). The left figures have gas (in blue)
         and one dust species; the right figures have gas (blue) and three dust
         species. The open circles represent the results from the
         \textsc{Phantom} simulation. The solid lines represent the analytical
         solution from \citet{Benitez-Llambay2019ApJS..241...25B}.%
         \label{fig:dustyshock_final}}
   \end{center}
\end{figure*}

\begin{figure*}
   \begin{center}
      \includegraphics[width=\textwidth]{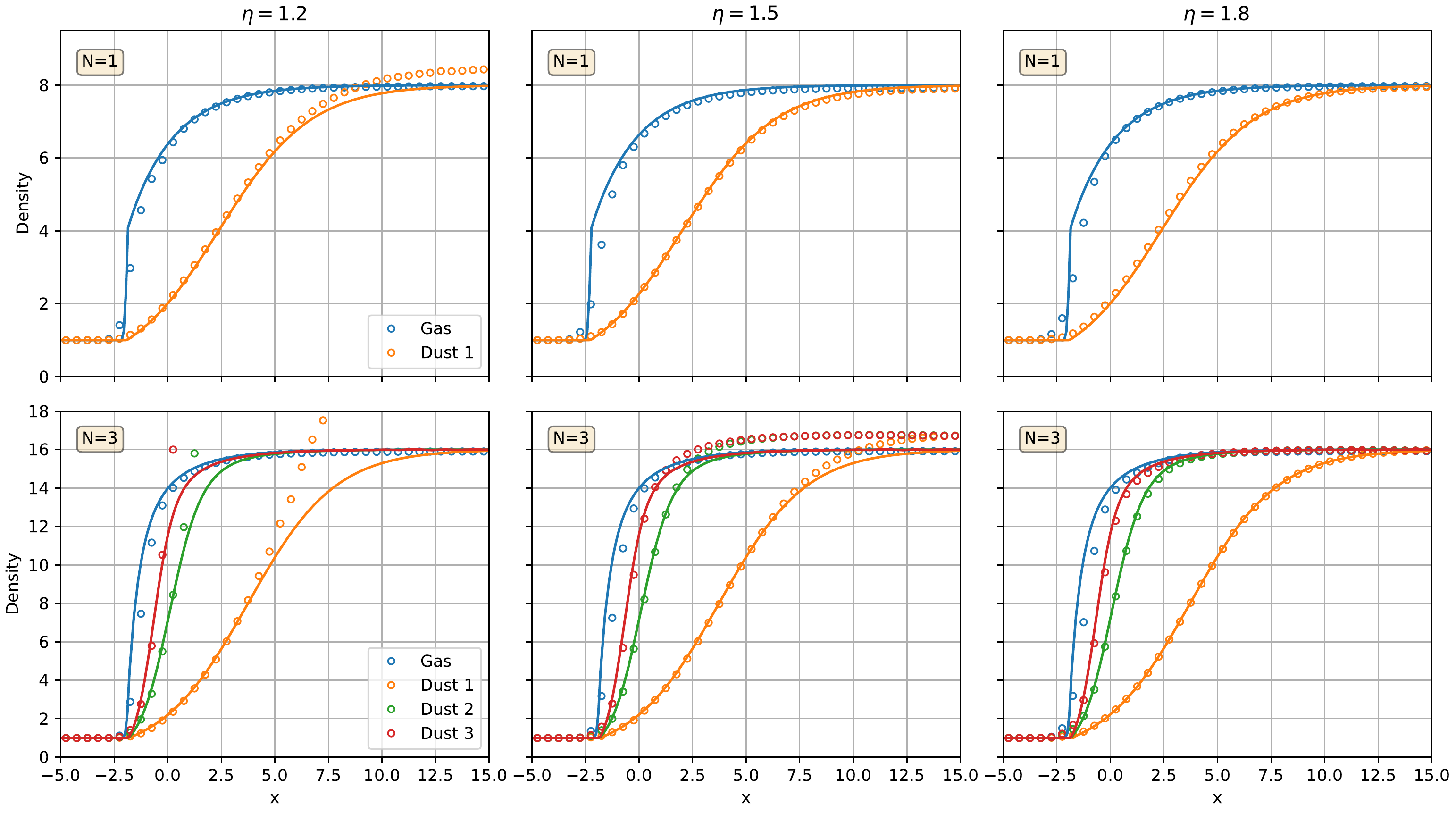}
      \caption{The effect of \(\eta\) in the dusty shock test problem.
         The lines and markers are the same as used in
         Figure~\ref{fig:dustyshock_final}. From left to right: \(\eta =
         1.2, 1.5, 1.8\), corresponding to \(\overline{N}_{\rm neigh} = 195,
         381, 660\). For the single dust species case we require 381 neighbours.
         For the three dust species case we require 660 neighbours.%
         \label{fig:dustyshock_hfact}}
   \end{center}
\end{figure*}

We performed a multiple species version of the isothermal steady state single
species dusty shock test, with exact solution given in
\citet{Lehmann2018MNRAS.476.3185L}.\footnote{The solution is described as a
J-type shock in which the fluid variables undergo a jump discontinuity, rather
than the properties being continuous across the front (as in a C-type shock).}
\citet{Benitez-Llambay2019ApJS..241...25B} generalised that solution to multiple
dust species. The solution, in steady state, for this shock is described by the
following set of equations:
\begin{align}
   \frac{\partial}{\partial x} \left( \rho_{\g} v_{\g} \right)
      &= 0, \label{eqn:shock-gas-continuity} \\
   \frac{\partial}{\partial x} \left( \rho_{\dd_i} v_{\dd_i} \right)
      &= 0, \label{eqn:shock-dust-continuity} \\
   \frac{\partial}{\partial x} \left[ \rho_{\g} (v_{\g}^2 + c_s^2)\right]
      &= - \sum_{i=1}^N K_i (v_{\g} - v_{\dd_i}), \\
   \frac{\partial}{\partial x} \left( \rho_{\dd_i} v_{\dd_i}^2 \right)
      &= K_i (v_{\g} - v_{\dd_i}).
\end{align}
We assume a constant drag coefficient \(K_i\) per species. The constant
isothermal sound speed is \(c_s\). In steady state, due to the drag force, the
gas and all dust species have the same pre-shock velocity \(v_s\) and the same
asymptotic post-shock velocity. We define normalised velocities \(\omega_{\g} =
v_{\g} / v_s\) and \(\omega_{\dd_i} = v_{\dd_i} / v_s\), then these equations
can be integrated to give
\begin{align}
   0 &= \omega_{\g}^2 + \omega_{\g} \left[ \sum_i \varepsilon_i (\omega_{\dd_i} - 1)
      - \mathcal{M}^{-2} -1 \right] + \mathcal{M}^{-2}, \\
   \frac{\dd \omega_{\dd_i}}{\dd x} &= \frac{K_i}{\rho_{\dd_i}}
      \left( \omega_{\g} - \omega_{\dd_i} \right),
\end{align}
where \(x_0\) is the shock position and \(\mathcal{M} = v_s / c_s\) is the Mach
number \citep{Benitez-Llambay2019ApJS..241...25B}. These equations can be
numerically integrated to give the normalised velocities. Then
Equations~\ref{eqn:shock-gas-continuity}~\&~\ref{eqn:shock-dust-continuity} give
the densities:
\begin{align}
   \rho_j &= \frac{\rho_j(x_0) v_j(x_0)}{v_j},
\end{align}
where the index \(j\) refers to either \(\g\) or \(\dd_i\).

\begin{table}
   \centering
   \begin{tabular}{cccccccc}
      \hline
      \hline
      Fluids & \(K_1\) & \(K_2\) & \(K_3\) & \(\rho^-\) & \(\rho^+\) & \(\omega^-\) & \(\omega^+\) \\
      \hline
      2 & 1.0 & - & - & 1.0 & 8.0 & 1.0 & 0.125 \\
      4 & 1.0 & 3.0 & 5.0 & 1.0 & 16.0 & 1.0 & 0.0625 \\
      \hline
   \end{tabular}
   \caption{Dusty shock parameters. Same as \citet{Benitez-Llambay2019ApJS..241...25B}.}%
   \label{tab:shock}
\end{table}

We set up two tests. One test with a single dust species to validate against
previous single dust species tests. And another test with three dust species to
validate our multiple dust species method. We used that same parameters as shown
in Table~3 in \citet{Benitez-Llambay2019ApJS..241...25B} reproduced here in
Table~\ref{tab:shock} for completeness. For both tests we chose \(\mathcal{M} =
2\), \(c_s = 1\), and \(\varepsilon_i = 1\) for all dust species. The left
(pre-shock) velocities are set be the Mach number, i.e. \(v_j= 2\), where \(j\)
represents both gas and all dust species. We set the left (pre-shock) gas
density \(\rho_{\g}\) to 1. We set the drag coefficients \(K_i\) to: 1.0 for the
single dust species test, and 1.0, 3.0, and 5.0 for the three dust species test.
The right (post-shock) asymptotic values are given by
\begin{align}
   \rho_{j}^{\rm R} &=
      \frac{\rho_{j}^{\rm L} \omega_{j}^{\rm L}} {\omega_{j}^{\rm R}}, \\
   \omega_j^{\rm R} &= \left(1 + N \right)^{-1} \mathcal{M}^{-2},
\end{align}
where \(j\) represents both gas and all dust species, \(N\) is the number of
dust species, and the superscripts L and R represent left and right asymptotic
states, respectively. We set \(\alpha_{\rm AV}\) to 1.

We set up initial conditions such that the velocities are constant with their
asymptotic values on either side of the shock at \(x_{\rm shock} = 0\). We set
up the densities similarly except that we smoothed the shock at shock boundary,
using a logistic function, i.e.
\begin{align}
   \rho(x) = \frac{\rho_{\rm L} e^{-kx} + \rho_{\rm R}}{1 + e^{-kx}},
\end{align}
where \(\rho_{\rm L}\) and \(\rho_{\rm R}\) are the density on the left and
right of the shock, respectively, and \(k\) is a factor determining the width of
the smoothing. We set \(k^{-1}\) to \(2 x_{\rm sep}\), where \(x_{\rm sep} =
0.1953125\) is the (resolution-dependent) separation between particles in the
x-direction in the low density region. This is due to the large (in principle,
infinite) gradient in density at the shock boundary leading to an
over-concentration of dust in the post-shock region.

Figure~\ref{fig:dustyshock_initial} shows a sample of the initial particle
positions (bottom) with the analytical density profile (top). For the gas and
each dust species, we set up the particles on two close-packed lattices; one on
each side of the shock, with the density and velocity set by their asymptotic
values. We use the same resolution for the gas and each dust species. Due to the
resolution following the mass in SPH, the numerical resolution is higher in the
high density region.

Figure~\ref{fig:dustyshock_final} shows the velocity and density of both tests
at \(t = 300\) which is long enough so that any transient behaviour has died
out. We have binned the particles into 40 bins in the x-direction and calculated
a mass-weighted average for both density and velocity. We have over-plotted the
analytical solution for comparison. Note that the shock position has drifted
from its initial position, perhaps, due to numerical dissipation. We have
shifted the analytical solution shock position to minimise the error, as in
\citet{Benitez-Llambay2019ApJS..241...25B}. Note that this drift has velocity
\(\sim 0.5\%\) of the pre-shock velocity, and \(\sim 5\%\) of the post-shock
velocity. Our numerical method reproduces the exact solution in this non-linear
test problem.

To achieve an accurate solution we used a larger number of particle neighbours
in the SPH density sum than is typical in dust-gas simulations with
\textsc{Phantom}. The typical mean neighbour number when using the quintic
kernel, used by default for dust-gas simulations, in calculating density sums is
113 \citep{Price2018PASA...35...31P}. The mean neighbour number
\(\overline{N}_{\rm neigh}\) is related to the proportionality constant \(\eta\)
by
\begin{align}
   \overline{N}_{\rm neigh} = \frac{4}{3} \pi
      {\left( R_{\rm kern} \eta \right)}^3.
\end{align}
For the quintic kernel, \( R_{\rm kern} = 3.0 \). So, by default, \(\eta =
1.0\). We used a \(\eta\) value of 1.8 in Figure~\ref{fig:dustyshock_final}.

Figure~\ref{fig:dustyshock_hfact} shows the comparison between the analytical
and numerical solutions for the density varying the value of \(\eta = 1.2, 1.5,
1.8\), corresponding to \(\overline{N}_{\rm neigh} = 195, 381, 660\). We can see
that the accuracy of the gas density (blue line and markers) is independent (or
weakly dependent) on the neighbour number. In contrast, the dust density depends
strongly on the neighbour number. To avoid large errors, we require \(\eta =
1.5\), for the single dust species case, and \(\eta = 1.8\), for the three dust
species case. We discuss the implications of this in
Section~\ref{sec:discussion} below.

\subsection{Radial drift}%
\label{subsec:radialdrift}

\begin{figure}
   \begin{center}
      \includegraphics[width=\columnwidth]{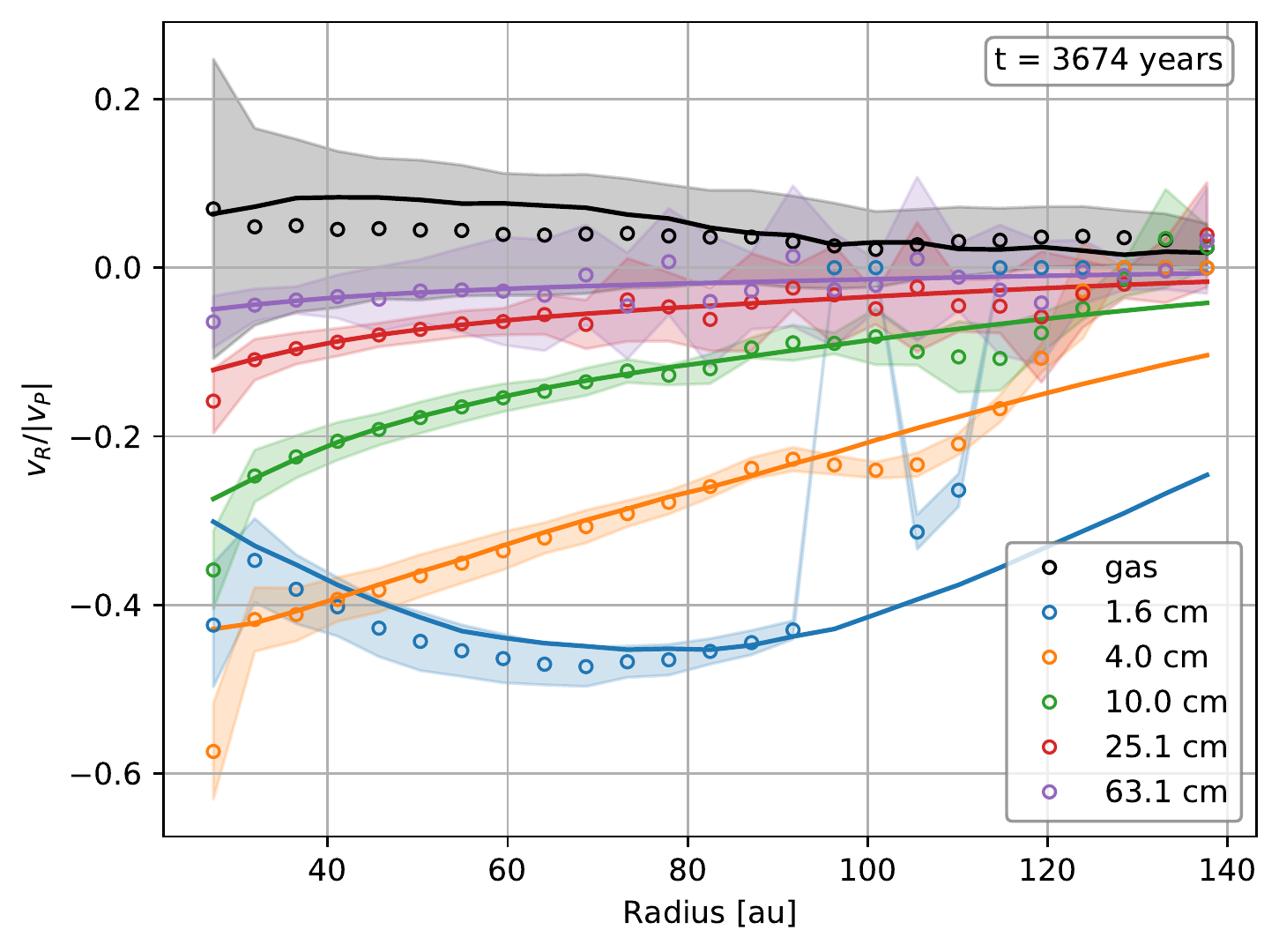}
      \caption{Azimuthally-averaged radial drift velocity profiles for gas
      (black) and five dust species (other colours). The circle markers
      represent the drift velocities measured directly on the particles with
      shading indicating the standard deviation from radial binning. The solid
      line corresponds to the analytical solution from
      \citet{Dipierro2018MNRAS.479.4187D}. The numerical and analytical
      solutions agree within the variation due to binning. The error in the
      1.6~cm and 4.0~cm grains for radius \(\gtrsim\)~90~au is due to the
      depletion of those grains at those radii.%
      \label{fig:radialdrift}}
   \end{center}
\end{figure}

We tested the method against the steady-state radial drift solution for a
viscous protoplanetary disc with multiple dust species calculated in
\citet{Dipierro2018MNRAS.479.4187D} as a generalisation of the single species
solution in \citet{Nakagawa1986Icar...67..375N} and the multiple species
inviscid case in \citet{Bai2010ApJ...722.1437B}. Following
\citet{Dipierro2018MNRAS.479.4187D}, the gas radial drift velocity \(v^{\rm
g}_R\) is given by
\begin{align}
   \label{eqn:radialdrift-gas}
   v^{\rm g}_R = \frac{-\lambda_1 v_{\rm P} + \left(1 + \lambda_0\right) v_{\rm visc}}
      {\left(1 + \lambda_0\right)^2 + \lambda_1^2},
\end{align}
and the dust radial drift velocity \(v^{\rm d_i}_R\) is given by
\begin{align}
   \label{eqn:radialdrift-dust}
   v^{\rm d_i}_R = \frac{v_{\rm P} \left[\left(1 + \lambda_0\right) \mathrm{St}_i - \lambda_1\right]
      + v_{\rm visc} \left(1 + \lambda_0 + \mathrm{St}_i \lambda_1\right)}
      {\left[\left(1 + \lambda_0\right)^2 + \lambda_1^2\right] \left(1 + \mathrm{St}_i^2\right)},
\end{align}
where \(v_{\rm P}\), the dust drift velocity due to the gas being
pressure-supported, is
\begin{align}
   \label{eqn:radialdrift-pressure}
   v_{\rm P} = \frac{1}{\rho_g \Omega_k} \frac{\partial P}{\partial R},
\end{align}
and \(v_{\rm visc}\), the gas radial velocity due to viscous spreading, is
\begin{align}
   \label{eqn:radialdrift-visc}
   v_{\rm visc} = \left[R \rho_g \frac{\partial}{\partial R} \left(R^2 \Omega_k\right) \right]^{-1}
      \frac{\partial}{\partial R} \left( \eta R^3 \frac{\partial \Omega_k}{\partial R} \right),
\end{align}
with \(\eta = \nu \rho_g = \alpha c_s H \rho_g\) (where \(\alpha\) is the
\citet{Shakura1973A&A....24..337S} viscosity parameter) and \(\lambda_k\) is
given by
\begin{align}
   \label{eqn:radialdrift-lambda}
   \lambda_k = \sum_i \frac{\mathrm{St}^k_i}{1 + \mathrm{St}_i^2} \epsilon_i.
\end{align}
We used Equations~\ref{eqn:radialdrift-gas}~\&~\ref{eqn:radialdrift-dust} to
validate our numerical method for the radial drift test.

We set up a 3-dimensional, locally isothermal protoplanetary disc around a star
of \(1~M_{\odot}\). We used 10 dust species logarithmically distributed in size
from 1~cm to \(10^4\)~cm. The gas mass was \(0.05~M_{\odot}\) and the total
dust-to-gas ratio was 0.5. The dust sub-disc mass for each species scaled with
the grain size. We used \(10^6\) gas particles and \(10^6\) dust particles with
\(10^5\) per species. We used a dust grain material density of 3~g/cm\({}^3\).
The disc extends from 1~au to 150~au. The radial profile of the surface density
was given by \(\Sigma(R) \propto R^{-1}\). The radial profile for the gas
temperature was given by \(T(R) \propto R^{-0.5}\) with the aspect ratio \(H/R =
0.05\) at \(R=1\)~au. We used \(\alpha\)-disc viscosity
\citep[Section~3.3.4]{Price2018PASA...35...31P} with an artificial viscosity
parameter \(\alpha_{\rm AV}=0.74\). We set the initial radial velocities to
zero. We set the orbital velocities to account for the pressure gradient
\citep[Section~3.3.2]{Price2018PASA...35...31P}. Initially, the dust sub-discs
are co-located with the gas disc. We used individual timestepping
\citep[Section~2.3.4]{Price2018PASA...35...31P}.

After evolving the disc for 3674 years, we binned the particles into 25 equally
spaced radial bins between 25~au and 140~au. For the analysis, we considered
only particles within 0.05 times the gas scale height of the midplane. The
Stokes number is exponentially dependent on the height above the midplane
(\(\mathrm{St_i} \sim \rho_g^{-1} \sim e^{z^2}\)). Averaging, within a radial
bin, by including particles above the midplane (\(\gtrsim 0.05\ H_{\rm g}\))
produces a poor estimate of the midplane Stokes number. We computed the
mass-weighted average of the radial velocities in the each annular cylinder for
the gas and each dust species. To compare with the analytical solution
(Equations~\ref{eqn:radialdrift-gas}~\&~\ref{eqn:radialdrift-dust}) we computed
each of the terms
(Equations~\ref{eqn:radialdrift-pressure}--\ref{eqn:radialdrift-lambda}) on the
radial bins. We computed the gradient terms using a second-order finite
difference approximation.

Figure~\ref{fig:radialdrift} shows this comparison of the radial drift
velocities for the 5 smallest dust species. The drift velocities are scaled by
\(v_{\rm P}\). We omit grains larger than 63.1~cm as they are consistent with
zero radial drift. The shading represents the standard deviation from the
binning procedure. The solid lines represent the drift velocities computed from
Equations~\ref{eqn:radialdrift-gas}~\&~\ref{eqn:radialdrift-dust}. We see that
for most of the radial extent of the disc the drift velocities match to within
the error introduced by binning. The deviation for the 1.6~cm and 4~cm grains
for radius \(\gtrsim\)~90~au is due to depletion of those grains via radial
drift.

\section{Discussion}%
\label{sec:discussion}

In deriving our method we made no assumptions on the size of the dust grains.
This is unlike the method described in \citet{Hutchison2018MNRAS.476.2186H}
which makes the terminal velocity approximation, appropriate for small,
well-coupled grains \citep{Youdin2005ApJ...620..459Y}.

However, we use an explicit timestepping scheme
\citep{Price2018PASA...35...31P}. In addition to the usual
Courant–Friedrichs–Lewy (CFL) time step constraint
\citep{Courant1928MatAn.100...32C}, which in SPH form is \(\Delta t_{\rm CFL} <
C_{\rm Cour} h/c_s\) \citep{Price2018PASA...35...31P}, there is an extra
constraint given in Equation~\ref{eqn:timestep-dust}. This restricts the minimum
stopping time of grains that can be simulated. To get an estimate on this
restriction we compare this constraint with the CFL constraint. If we require
that the stopping time constrained time step is less restrictive than the CFL
constraint then we get that
\begin{align}
   t_{s_i} > \frac{C_{\rm Cour} h (1 - \varepsilon)}{C_{\rm drag} c_s}.
\end{align}
The parameters \(c_s\) and \(\varepsilon\) are physical parameters set by the
problem of interest. Whereas \(C_{\rm Cour} = 0.3\), \(C_{\rm drag} = 0.9\), and
\(h\) are numerical parameters.

Considering protoplanetary discs, converting this into an inequality for the
midplane Stokes number \(\mathrm{St}_i\) we get
\begin{align}
   \mathrm{St}_i > \frac{C_{\rm Cour}}{C_{\rm drag}} (1 - \varepsilon) \frac{h}{H},
\end{align}
where \(H\) is the gas disc scale height. Using values of \(h/H \approx 0.5\)
and \(\varepsilon \approx 0.1\) we find a minimum Stokes number, such that the
time step is unconstrained compared with the CFL condition, of \(\mathrm{St}_i
\gtrsim 0.1\). Note that the numerical resolution is inversely proportional to
\(h\). So, as the resolution increases this restriction is loosened. However, as
the resolution increases the CFL timestep decreases.

The Stokes number is linearly proportional to the grain size (in the Epstein
regime). So this constraint on the Stokes number provides a constraint on the
dust grain size; one that is dependent on the disc parameters too. Considering a
disc with surface density \(\Sigma \approx 1\)~g/cm\({}^2\) and effective grain
size \(\varrho_{\rm eff} = 3\)~g/cm\({}^3\) we find that the minimum grain size
is approximately 0.2~mm. Note that this is \emph{not} a minimum grain size that
can be represented. Using smaller grains is merely inefficient given the
explicit time stepping scheme. The single fluid method is more efficient for
these smaller grains, which satisfy the terminal velocity approximation.
\citet{Cuello2019MNRAS.483.4114C} show that the two methods overlap and give the
same results. Using the single fluid method for small grains negates the need to
implement an implicit timestepping scheme as in, e.g.,
\citet{Loren-Aguilar2014MNRAS.443..927L, Loren-Aguilar2015MNRAS.454.4114L}.

The method requires storing an extra set of particles per dust species. For each
particle we store the position, smoothing length, and velocity, in total
requiring seven 64-bit floating point numbers per particle. This is in contrast
to the mixture method. In that method there is only a single set of particles
(representing the mixture) and for each dust species an additional four 64-bit
floating point numbers are required: one for the dust fraction, and three for
the velocity differential \citep{Hutchison2018MNRAS.476.2186H}. So it seems that
the method described here is more expensive in memory and storage requirements
than \citet{Hutchison2018MNRAS.476.2186H}. However, this is not so. In practice,
we use fewer dust SPH particles than gas; see, for example,
\citet{Dipierro2015MNRAS.453L..73D, Mentiplay2019MNRAS.484L.130M,
Calcino2019MNRAS.490.2579C}. This is to prevent the dust from becoming trapped
under the gas resolution scale \citep{Laibe2012MNRAS.420.2345L}. In either case
the additional memory and storage requirement is not prohibitive.

We represent each dust species as a pressureless fluid that interacts with gas
via the drag force and with massive objects such as stars and planets
(represented by sink particles) by the gravitational force. The gas, in
contrast, has a non-zero pressure. The pressure gradient force leads to a
rearrangement of particles into a regular glass-like lattice
\citep{Monaghan2005RPPh...68.1703M} that minimises the Lagrangian of the system
of particles \citep{Price2012JCoPh.231..759P}. So, if the gas particles are
initialised in a ``pathological'' arrangement, e.g.\@ such that particles are
almost on top of each other or randomised, they will rearrange into a
quasi-regular lattice. This rearrangement leads to an improved density estimate
\citep{Price2012JCoPh.231..759P}. The lack of rearranging due to the lack of a
pressure gradient force in the dust can lead to noisy density estimates.

In calculating the density we sum over neighbours of the same type. So the gas
and each dust species has a smoothing length per particle independent of each
other species. In \textsc{Phantom}, the smoothing length proportionality factor
\(\eta\) is constant. This means that the mean number of neighbours
\(\overline{N}_{\rm neigh}\) is the same for both the gas and dust. Given that
the dust particle arrangement does not re-mesh, unlike the gas, using a larger
number of neighbours may improve the density estimate for the dust. One possible
approach, not explored in this study, would be to set \(\eta\) independently for
the gas and dust. The dusty shock test (Section~\ref{subsec:shock}) provides
some evidence for this. In that test the gas density was accurate for \(\eta =
1.2\). However, the dust density required \(\eta = 1.8\) in the three dust
species case.

An alternative approach would be to provide a pressure gradient force for the
dust. This would provide for a rearrangement of particles to improve the density
estimates and reduce artificial clumping of dust. Such a force would necessarily
be a short range one. It is not clear at what distance it should activate. Any
such prescription would be a sub-grid model and resolution dependent.

In the limit of small drag coefficient, or large stopping time, our scheme
reduces to an N-body integrator, which preserves the orbital parameters exactly.
As the method is Lagrangian, we can track particles throughout time. This allows
for coupling with, for example, chemistry codes that require knowledge of the
thermal history of the dust.

\section{Conclusions}

We have derived a smoothed particle hydrodynamics numerical scheme to model a
dust and gas mixture using a separate set of SPH particles per species. The
method includes a drag force coupling between the gas and each dust species that
conserves momentum. Thus the method captures the full effects of backreaction
between the gas and dust. This method can be applied to any distribution of dust
grain sizes. It is not restricted to grain sizes such that the terminal velocity
approximation holds. Although, given that it uses an explicit time stepping
scheme, it becomes inefficient for grain sizes with small Stokes number, i.e.
\(\lesssim 0.05\).

We have implemented this method in the SPH code \textsc{Phantom}. We have
demonstrated that the method is accurate by testing it on four test problems
with analytical solutions: a dusty box, to test the drag coupling and time
stepping in the absence of spatial gradients; a dusty wave, to test those
factors with spatial gradients in a linear regime; a dusty shock, to test those
factors in a highly non-linear regime; and a radial drift test to test the
method is the context of a global 3-dimensional protoplanetary disc model. In
all cases we show the method is accurate.

We suggest that dust particles may require a larger number of particle
neighbours in computing SPH sums than the gas. This is due to the lack of
inter-particle forces, i.e.\ dust is pressureless fluid and does not rearrange
into a glassy structure like the gas does. Dust can numerically clump resulting
in artificial high density regions. Using a greater number of particle
neighbours leads to a smoother density field. This may reduce the performance
degrading effect of high density dust clumps on the time step.

\section*{Acknowledgements}

DM is funded by a Research Training Program Stipend from the Australian
government. We acknowledge Australia Research Council funding via DP180104235,
FT130100034, and FT170100040. GL acknowledges funding from the European Union's
Horizon 2020 research and innovation program under the Marie Sk\l{}odowska-Curie
grant agreement No 823823 and ERC CoG project PODCAST No 864965. We used OzStar,
funded by Swinburne University of Technology and the Australian government, for
computation. DP thanks Pablo Benítez-Llambay for useful discussions at the Great
Barriers in Planet Formation conference, 2019. We used \textsc{Phantom} to
perform the numerical simulations \citep{Price2018PASA...35...31P}. We used
\textsc{Plonk} for analysis and visualisation of \textsc{Phantom} data
\citep{Mentiplay2019JOSS....4.1884M}. We also used the following Python packages
and tools: \textsc{NumPy} \citep{Oliphant2006, van-der-Walt2011CSE....13b..22V},
\textsc{h5py} \citep{Collette2013}, \textsc{matplotlib}
\citep{Hunter2007CSE.....9...90H}, \textsc{pandas} \citep{McKinney2010},
\textsc{IPython} \citep{Perez2007CSE.....9c..21P}, and \textsc{Jupyter}
\citep{Kluyver2016}.

\section*{Data availability}

The data underlying this article can be reproduced following instructions in a
file \texttt{README.md} in the repository ``multigrain'' (available on GitHub at
\url{https://github.com/dmentipl/multigrain} and archived on Figshare at
\url{https://doi.org/10.6084/m9.figshare.13076435}). This repository contains
Python code used to set up the tests and generate data, analyse the output data
from the tests, and produce the figures of the manuscript. It contains a file
\texttt{environment.yml} that can be used---along with
\href{https://conda.io/}{conda}, a cross-platform, software package manager---to
set up a computing environment with the required software dependencies. We used
the Phantom version with git hash
\href{https://github.com/danieljprice/phantom/commit/64dbd2b124ca74051eed920d6cad0a2e83157478}{\texttt{64dbd2b1}}
with some patches described in the ``multigrain'' repository.




\bibliographystyle{mnras}
\bibliography{multigrain-paper}





\bsp 
\label{lastpage}
\end{document}